\documentclass[prd,onecolumn,notitlepage,eqsecnum,nofootinbib,floatfix]{revtex4-1}
\usepackage{amssymb}
\usepackage{amsmath}
\usepackage{amsfonts} 
\usepackage{bm}
\usepackage{color} 
\usepackage{graphicx}
\usepackage[caption=false]{subfig}
\newcommand{\stf}[1]{{\langle #1 \rangle}} 
\renewcommand{\Re}{\mbox{Re}} 
\renewcommand{\Im}{\mbox{Im}} 
\newcommand{\m}{\mathfrak{m}} 
\newcommand{\f}{\mathfrak{f}} 
\allowdisplaybreaks
\begin{document}
\title{Self-force and fluid resonances}   
\author{Soichiro Isoyama} 
\affiliation{Department of Physics, University of Guelph, Guelph,
  Ontario, N1G 2W1, Canada} 
\author{Raissa F.~P.~Mendes} 
\affiliation{Department of Physics, University of Guelph, Guelph,
  Ontario, N1G 2W1, Canada} 
\author{Eric Poisson} 
\affiliation{Department of Physics, University of Guelph, Guelph,
  Ontario, N1G 2W1, Canada {\ }} 
\affiliation{Canadian Institute for Theoretical Astrophysics,
  University of Toronto, Toronto, Ontario, M5S 3H8, Canada} 
\date{\today}
\begin{abstract} 
The gravitational self-force acting on a particle orbiting a massive
central body has thus far been computed for vacuum spacetimes
involving a black hole. In this work we continue an ongoing effort to
study the self-force in nonvacuum situations. We replace the black
hole by a material body consisting of a perfect fluid, and determine
the impact of the fluid's dynamics on the self-force and resulting
orbital evolution. We show that as the particle inspirals toward the
fluid body, its gravitational perturbations trigger a number of
quasinormal modes of the fluid-gravity system, which produce resonant
features in the conservative and dissipative components of the
self-force. As a proof-of-principle, we demonstrate this phenomenon in
a simplified framework in which gravity is mediated by a scalar
potential satisfying a wave equation in Minkowski spacetime.     
\end{abstract} 
\maketitle

\section{Introduction and summary} 
\label{sec:intro} 

The inspiral of a solar-mass compact object into a supermassive black 
hole is a promising source of low-frequency gravitational waves 
for space-based detectors such as eLISA \cite{elisa:13} and DECIGO
\cite{decigo:11}. This observation has motivated a sustained effort 
to model the orbital motion of such a system, in a treatment that
goes beyond the test-mass approximation in which the compact 
object moves on a geodesic in the background spacetime of the
large black hole. In this improved treatment \cite{mino-etal:97,
  quinn-wald:97}, the gravitational influence of the small body is
taken into account, and the motion is geodesic in the perturbed
spacetime instead of the background spacetime. Alternatively, the
motion of the small body can still be viewed in the background
spacetime, where it is now accelerated, the acceleration being caused
by the gravitational perturbation created by the body. In this view,
the body experiences a gravitational self-force, and it is this
self-force that is responsible for the inspiraling motion.  

To date the gravitational self-force was formulated rigorously 
\cite{gralla-wald:08, pound:10a, pound:10b}, it was computed and
implemented in orbital evolutions around nonrotating black holes 
\cite{warburton-etal:12, osburn:15}, it was implicated in an improved
calculation of the innermost circular orbit of a Kerr black hole
\cite{isoyama-etal:14}, and its role was elucidated in attempts to
violate cosmic censorship by overspinning a near-extremal black hole
\cite{jacobson-sotiriou:09, barausse-cardoso-khanna:10,
  colleoni-barack:15, colleoni-etal:15}.  
The gravitational self-force was extended to second order in
perturbation theory \cite{detweiler:12, gralla:12, pound:12a, 
  pound:12b, pound-miller:14}, and its consequences were compared
to the predictions of high-order post-Newtonian theory
\cite{shah-friedman-whiting:14, johnsonmcdaniel-shaw-whiting:15,
  ackay-etal:15} and numerical relativity 
\cite{letiec-etal:13}. Other achievements of the gravitational
self-force program are reviewed in Refs.~\cite{barack:09,
  poisson-pound-vega:11}.   

Until recently the gravitational self-force was formulated and
computed for bodies moving in vacuum spacetimes, for which the
background metric satisfies the Einstein field equations in the
absence of a matter term. This restriction, of course, is amply
justified for the applications reviewed previously, in which the
large body is a black hole. Other applications, however, may well
involve the presence of matter. An example of such a situation is 
the recent effort \cite{zimmerman-etal:13} to take into account 
self-force effects in attempts to overcharge a near-extremal black 
hole \cite{hubeny:99} (this is the charged version of the
overspinning scenario described previously). In this situation 
the black hole is charged, the spacetime is filled with an
electrostatic field, and the coupling between gravitational and
electromagnetic perturbations creates complications in the
formulation of the self-force. In this case an adequate formulation
of the self-force for background spacetimes containing either a
scalar or electromagnetic field was provided by Zimmerman and 
Poisson \cite{zimmerman-poisson:14, zimmerman:15a} and
Linz, Friedman, and Wiseman \cite{linz-friedman-wiseman:14}. In a
related development, the self-force was extended to scalar-tensor
theories of gravity by Zimmerman \cite{zimmerman:15b}.  

The presence of a scalar or electromagnetic field in the background
spacetime makes a concrete and convenient starting point for the
formulation of the self-force in nonvacuum situations, but other
applications call for the presence of other types of matter
fields. The example that motivates the work presented in this paper is
the self-forced motion of a small object around a large material body
made up of a perfect fluid. There is currently no formulation of the
self-force that accounts for the coupling between the gravitational
and fluid perturbations that arise in such systems. A self-force
formulation that achieves this would have some use. For example,
one could use this self-force to test the hypothesis that the central
body is a black hole by comparing its predictions to those of an
alternative hypothesis, in which the central body is a fluid body in
hydrostatic equilibrium.  

Our goal in this paper is not to provide such a general formulation
of the self-force. What we wish to do instead is to explore the
expected physical consequences of the gravity-matter coupling on the 
self-force. This coupling gives rise to a discrete spectrum of 
quasinormal modes, and we wish to explore the impact of these modes 
on the gravitational self-force. What intrigues us the most is the
possibility that resonances can occur when the orbital frequency is in
a commensurate relation with a quasinormal-mode frequency; we wish to
determine the role of these resonances on the inspiraling motion of
the particle. The purely dissipative aspects of this problem are
already understood, and can be inferred from the impact of the
resonances on the radiative fluxes of energy and angular momentum
from the system \cite{kojima:87, gualtieri-etal:01, pons-etal:02}. 
But we wish here to consider all aspects of the self-force, both
conservative and dissipative. Our analysis is a rudimentary one that
serves as a proof-of-principle. We believe that our results motivate a
deeper investigation, which we leave for future work.   

In this spirit of proof-of-principle, we choose to avoid the technical
complications associated with a perturbed fluid system in general
relativity. For this exploration we consider a much simpler theory of 
gravity implicating a flat spacetime with Minkowski metric
$\eta_{\alpha\beta}$ and a scalar gravitational potential $\Phi$. 
(In this paper, we adopt geometric units, setting $G = c = 1$, and
metric signature $+2$. Latin and Greek indices run from 1 to 3, and 
from 0 to 3, respectively.)
The fluid is described by its energy-momentum tensor
$T^{\alpha\beta}$. The gravitational field equation is chosen to be  
\begin{equation} 
\Box \Phi = 4\pi T^\mu_{\ \mu}, 
\end{equation} 
and the fluid equations are 
\begin{equation} 
\nabla_\beta T^{\alpha\beta} = -T^\mu_{\ \mu} \nabla^\alpha \Phi, 
\end{equation} 
where $\nabla_\alpha$ is the covariant-derivative operator compatible
with the Minkowski metric. This special-relativistic theory is 
obviously incompatible with observational tests, and we certainly 
do not claim that it is a viable description of gravity. It is not. 
We find, nevertheless, that it provides a useful and simple framework 
to explore the impact of the fluid-gravity coupling on the self-force.
One important feature of the theory is that it automatically
incorporates a source of dissipation:
the gravitational potential satisfies a wave equation, and the
gravitational waves carry energy away from the system. This radiative
loss permits the inspiral of a particle moving around a fluid body. 
The full set of equations associated with the scalar theory is
presented in Sec.~\ref{sec:equations}. 

In Sec.~\ref{sec:star} we construct stellar models in the scalar
theory of gravity. We consider a static and spherically symmetric
distribution of perfect fluid, and choose the equation of state to be
a slight modification of $p \propto \rho^2$, where $p$ is the
pressure and $\rho$ the rest-mass density; the modified (but still
physical) equation of state allows us to find simple analytical
solutions to the structure equations. Our unperturbed body has a mass
$M$ and radius $R$. 

\begin{figure} 
\includegraphics[width=0.85\linewidth]{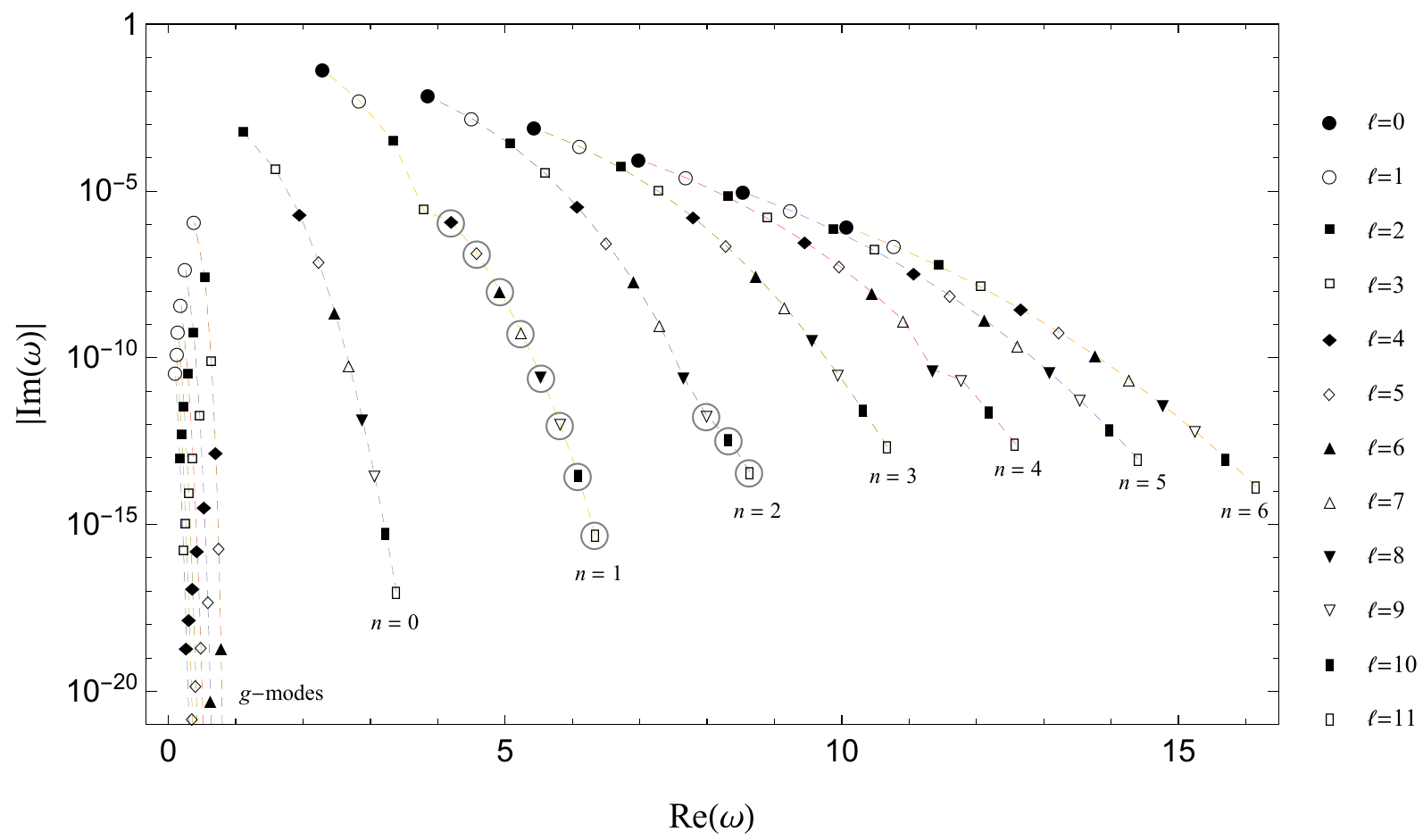}
\caption{Quasinormal modes of a fluid body with compactness 
  $M/R = 0.3$ in the scalar theory of gravity presented in the main
  text. The modes are characterized by the real and imaginary parts 
  of the eigenfrequency $\omega$, which are   presented in units of 
  $\sqrt{M/R^3}$. The modes are labelled by values of the
  multipole order $\ell$ and the number $|n|$ of nodes of $\xi_r$, the
  radial component of the Lagrangian displacement vector. The modes
  with $n=0$ are fundamental modes ($f$-modes), those with $n > 0$ are
  pressure modes ($p$-modes), and those with $n < 0$ are gravity modes 
  ($g$-modes). The modes enclosed by circles are unstable, with
  $\Im(\omega) > 0$.}  
\label{fig:fig1} 
\end{figure} 

In Sec.~\ref{sec:perturbed} we perturb the stellar model and derive
the coupled equations that govern the fluid and gravitational
perturbations. In the absence of an external forcing term, the equations
are homogeneous and represent an eigenvalue system for the
quasinormal modes. We integrate these equations, and obtain a number
of modes, those displayed in Fig.~\ref{fig:fig1}. The spectrum
delivered by the scalar theory is fairly typical, but we notice the
presence of some unstable modes (indicated with circles in the figure).
We also see that the mode lifetime (measured by $|\Im(\omega)|^{-1}$) 
can be extremely long for some modes, and this shall have an impact on
the width of the resonances identified below.  

\begin{figure} 
\includegraphics[width=1.0\linewidth]{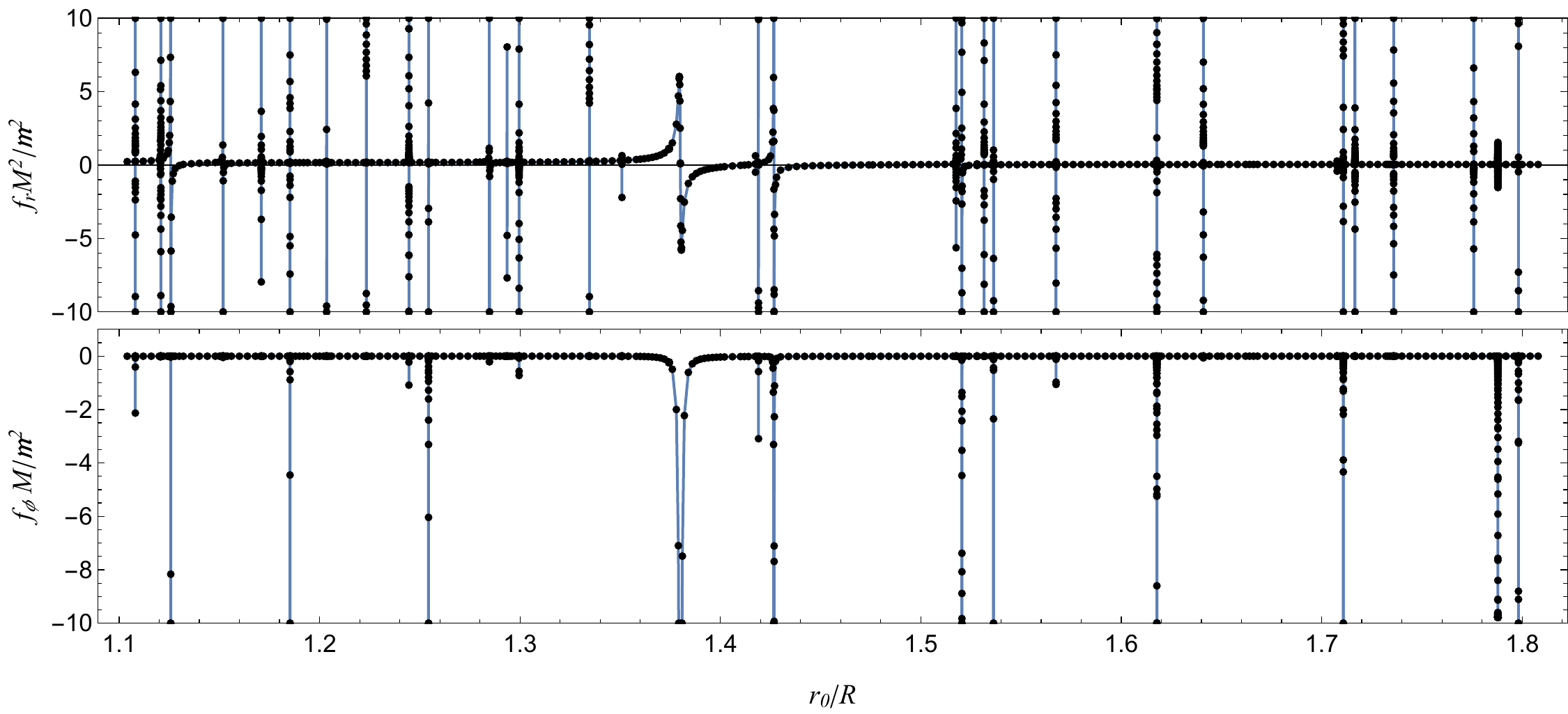}
\caption{Self-force on a particle of mass $\m$ placed on a circular
  orbit of radius $r_0$ around a star with compactness $M/R = 0.3$. 
  The radial component (upper panel) is
  presented in units of $\m^2/M^2$, and the angular component (lower
  panel) in units of $\m^2/M$. The orbital radius is displayed in
  units of the stellar radius $R$, for the range $1.10 < r_0/R <
  1.81$. We see multiple resonances producing large excursions of the
  self-force from its typical nonresonant value (which is of the order
  of $10^{-2}$ for $\f_r M^2/\m^2$, and $10^{-3}$ for 
  $\f_\phi M/\m^2$).}   
\label{fig:fig2} 
\end{figure} 

In Sec.~\ref{sec:particle} we place a particle of mass $\m$ on a
circular orbit of radius $r_0$ and angular frequency $\Omega$ around
the fluid star, and reformulate the perturbation equations to account
for the presence of the particle. The force exerted on the
particle by the star is $f_\alpha = \m (\delta_\alpha^{\ \beta} 
+ v_\alpha v^\beta) \nabla_\beta \Phi$, where $v^\alpha$ is the
particle's velocity vector, and the self-force is 
\begin{equation} 
\f_\alpha = \m \bigl( \delta_\alpha^{\ \beta}  
+ v_\alpha v^\beta \bigr) \nabla_\beta \delta \Phi, 
\end{equation} 
where $\delta \Phi$ is the perturbation created by the particle itself 
(which must be regularized before it is evaluated on the world line). 
For a circular orbit the self-force features two independent components. 
The first is $\f_\phi$, which is responsible for all dissipative aspects
and drives the particle's inspiral; the second is $\f_r$, which
provides a conservative correction to the relation between $\Omega$
and $r_0$. These components of the self-force are displayed in
Fig.~\ref{fig:fig2}. We see the resonances at work in the dramatic
excursions of the self-force from the typical nonresonant value. A
resonance is produced whenever $\Omega = \Re(\omega)/m$, where 
$\omega$ is the eigenfrequency of a quasinormal mode of multipole 
order $\ell$, and $m \leq \ell$ is an integer. Most resonances are
extremely narrow; this is a consequence of the long lifetimes of the
associated quasinormal modes. A notable exception is the resonance 
with the $\ell = 2$ fundamental mode of the fluid, at
$r_0/R \simeq 1.38$, which is broad. We were able to associate each
resonant feature in the self-force with a specific quasinormal mode;
our discussion in Sec.~\ref{sec:particle} provides these details. 

\begin{figure} 
\includegraphics[width=0.8\linewidth]{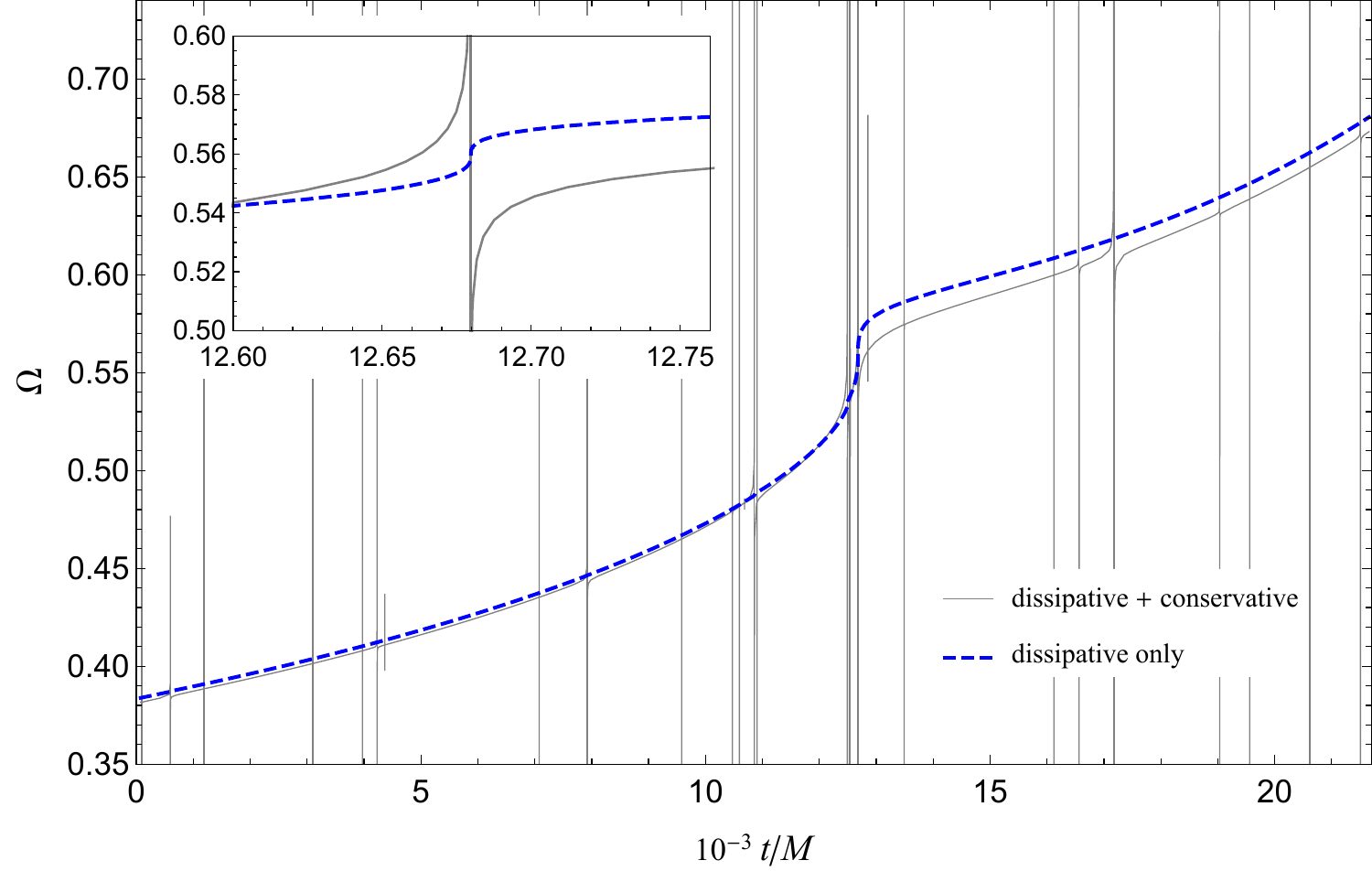}
\caption{Orbital angular velocity as a function of time resulting from
  the action of the self-force. The angular velocity is presented in
  units of $\sqrt{M/R^3}$, time is presented in units of $M$, and the
  orbital evolution was calculated for $\m/M = 0.01$. The particle
  is initially set at $r_0 = 1.80$ when $t=0$ and evolves until it
  reaches $r_0 = 1.20$. The dashed blue line represents an evolution 
  that incorporates only $\f_\phi$, the dissipative component of the
  self-force. The solid gray line represents an evolution calculated
  with both $\f_r$ and $\f_\phi$, and which therefore includes also the
  conservative component of the self-force. The inset highlights the
  interval around $t/M = 12,680$, when the particle crosses the
  broad resonance at $r_0/R \simeq 1.38$.}
\label{fig:fig3} 
\end{figure} 

\begin{figure} 
\includegraphics[width=0.8\linewidth]{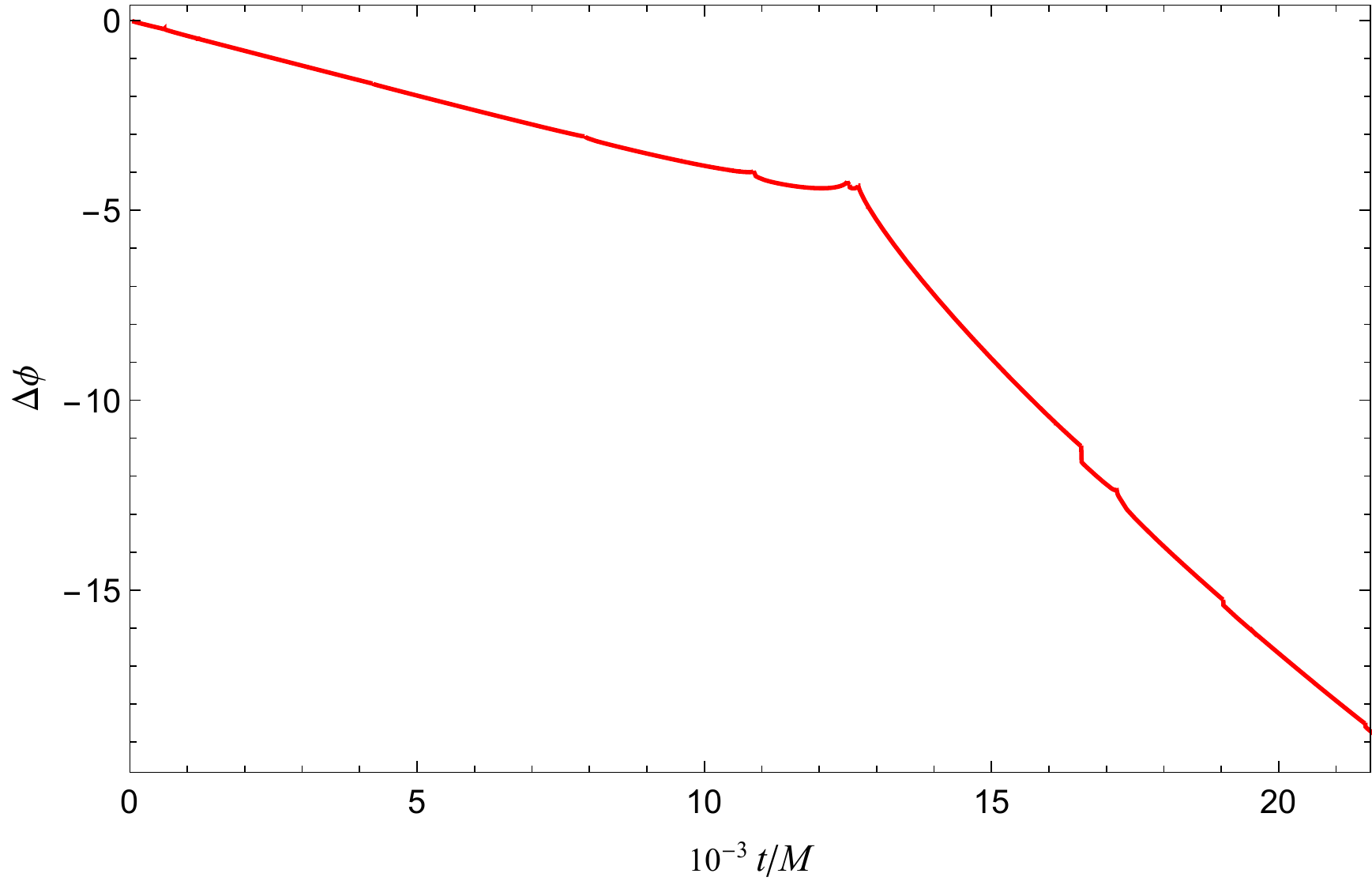}
\caption{Dephasing $\Delta \phi = \phi - \phi_0$ of the actual
  orbital evolution, given by $\phi(t)$, relative to the purely
  dissipative evolution, given by $\phi_0(t)$, as a function of time. 
  We see that after the broad resonance at $r_0/R \simeq 1.38$ is
  crossed (at $t/M = 12,680$), there is a perceptible increase in the
  rate of dephasing.} 
\label{fig:fig4} 
\end{figure}

The self-force is next incorporated (also in Sec.~\ref{sec:particle}) 
into the particle's equations of motion, in an approximate treatment
that assumes that the orbital evolution proceeds on a time scale that
is long compared with the orbital period. The impact of the self-force
can be seen in Fig.~\ref{fig:fig3}. The blue dashed line in the figure
represents an evolution that incorporates only $\f_\phi$, the
dissipative component of the self-force. We see that $\Omega$
increases with time, which reflects the decreases in orbital radius
that accompanies a radiative loss of orbital energy. The rate of
increase is much larger near $t/M \simeq 12,680$, and this corresponds 
to the particle crossing the broad resonance at $r_0/R \simeq
1.38$. The solid gray line represents an evolution calculated with both
$\f_\phi$ and $\f_r$, and which therefore incorporates also the
conservative aspects of the self-force. We see that $\f_r$ produces
large excursions around the purely dissipative evolution. As the
particle enters a resonance, the angular frequency increases extremely
rapidly, then decreases, and finally increases again to rejoin the
dissipative evolution. These features are extremely narrow, reflecting
again the long lifetimes of the associated quasinormal modes. An
exception occurs with the broad resonance at $r_0/R \simeq 1.38$,
where we see the same cycle of increase, decrease, and re-increase, but
where we don't see the full evolution rejoining the dissipative
evolution. We observe instead that the dissipative evolution 
over-estimates the angular frequency by a few percent as the particle
comes out of the resonance. 
The mismatch between the actual and dissipative angular frequencies
observed in Fig.~\ref{fig:fig3} creates a dephasing $\Delta \phi$
between the two evolutions. We plot this as a function of time in
Fig.~\ref{fig:fig4}, and the figure reveals that the dephasing indeed 
becomes significant over the inspiral time scale.

Except for the broad resonance near $r_0/R \simeq 1.38$, the resonant
features observed in the self-force and the resulting orbital
evolution are all extremely narrow, a consequence of the long
lifetimes of the associated quasinormal modes. Nevertheless, these
features can in principle be measured; in particular, the sudden
excursions of the angular frequency around the smooth, dissipative
curve would have a manifestation in the gravitational waves emitted by
the particle. In situations involving shorter lifetimes and broader 
resonances, it can be expected that the resonant features would 
reveal valuable information regarding the body's internal composition. 
We feel that our findings in this paper motivate a more thorough 
analysis carried out in full general relativity.  

The rest of the paper is devoted to a detailed derivation of the
results presented above, accompanied with a more complete discussion.
To set the stage, we consider in Sec.~\ref{sec:Newtonian} the
simplest model of a resonant self-force, implicating Newtonian gravity
and a fluid body with a uniform mass density (see also 
Refs.~\cite{lai:94, ho-lai:99} for related Newtonian analyses). 
In this language the self-force is the result of a dynamical
tidal interaction between the particle and the body. As we shall see, 
many of the features of our relativistic model are captured by this 
simplest of models.

\section{Newtonian model of resonant self-force} 
\label{sec:Newtonian} 

As stated, the resonant features described previously in the
scalar-gravity self-force can be captured by a strictly Newtonian
model. In this situation a particle of mass $\m \ll M$ raises a
dynamical tide on a fluid body of mass $M$ and radius $R$, and the
resulting deformation of the body's gravitational potential acts back
on the particle, thereby producing what can be described as a
self-force. For simplicity we take the unperturbed configuration to
be a spherical fluid distribution of uniform mass density, and we 
truncate the tidal interaction at the leading, quadrupole order. In 
this simplified model the tidal interaction is mediated entirely by the 
$f$-mode of the fluid's quadrupolar perturbation. The Newtonian physics 
of dynamical tides is reviewed in Sec.~2.5.3 of 
Ref.~\cite{poisson-will:14}, and we shall import results and notation 
from this reference.   

\subsection{Tidal dynamics} 

The dynamical tide is driven by the particle's gravity, which is
described in terms of the Newtonian potential $U_{\rm part}(t,\bm{x})$;
the spatial coordinates $\bm{x}$ measure a displacement away from the
body's center-of-mass. The tidal interaction can be developed as a
multipole expansion, and at the leading order, the tidal environment
is described by the tidal quadrupole-moment tensor 
${\cal E}_{ab}(t) := -\partial_{ab} U_{\rm part}(t,\bm{0})$, in which
the particle's potential is evaluated at $\bm{x} = \bm{0}$ after
differentiation. The body's tidal response is measured by $I_{ab}(t)$,
the quadrupole-moment tensor of its mass distribution (defined as a 
tracefree tensor). This is given by 
\begin{equation} 
I_{ab} = -\frac{2}{5} M R^2 {\cal F}_{ab}, 
\label{tidalresp1} 
\end{equation} 
where 
\begin{equation} 
{\cal F}_{ab}(t) := \frac{1}{\omega} \int_{-\infty}^t {\cal E}_{ab}(t') 
e^{-\kappa(t-t')} \sin \omega (t-t')\, dt' 
\label{tidalresp2} 
\end{equation} 
is the fluid's response function; the real part of the mode frequency
is denoted $\omega$, and its imaginary part is denoted $-\kappa$. The
discussion in Ref.~\cite{poisson-will:14} excludes dissipation, and as
a consequence, the exponential factor does not appear in their
Eq.~(2.290).  It was inserted here to incorporate a source of
dissipation within the system. In the relativistic model examined
below, dissipation is naturally exhibited and associated with the
emission of gravitational radiation; here it is inserted by hand, and
associated with viscosity acting within the body. For the quadrupole
$f$-mode of a fluid with uniform density, 
$\omega^2 = 4M/(5 R^3)$. We assume that $\kappa \ll \omega$ to 
reproduce the long lifetimes of the relativistic quasinormal modes; we
explicitly assume that $\kappa > 0$, so that the $f$-mode is stable.  

We take the particle to move on a circular orbit of radius $r$ and
angular velocity $\Omega = \sqrt{M/r^3}$ around
the fluid body. For this situation we have 
\begin{equation} 
{\cal E}_{ab} = -\frac{3 \m}{r^3} n_\stf{ab}, 
\label{tidalmoment} 
\end{equation}  
where $n_\stf{ab} := n_{a} n_{b} - \frac{1}{3} \delta_{ab}$ is a
symmetric tracefree tensor formed from  
\begin{equation} 
\bm{n} := ( \cos\phi, \sin\phi, 0), 
\end{equation} 
a unit vector that points from the body's center-of-mass to the
particle, at which $\phi = \Omega t$. The coordinate system is
oriented in such a way that the orbital motion takes place in the
$x$-$y$ plane. For subsequent calculations it is necessary to complete
the vectorial basis with  
\begin{equation} 
\bm{\lambda} := ( -\sin \phi, \cos \phi, 0 ); \qquad 
\bm{e} := (0,0,1). 
\end{equation} 
The unit vector $\bm{\lambda}$ points in the direction of the
particle's velocity, and $\bm{e}$ is normal to the orbital plane. 

Inserting  Eq.~(\ref{tidalmoment}) into Eq.~(\ref{tidalresp2}) and
performing the integration returns 
\begin{equation} 
{\cal F}_{ab} = \frac{15}{4} \frac{\m}{M} \biggl( \frac{R}{r}
\biggr)^3 \Bigl[ 2 A n_{(a} \lambda_{b)} 
-2 B n_\stf{a b} + (3C-B) e_\stf{a b} \Bigr], 
\label{F_expression} 
\end{equation} 
where 
\begin{subequations} 
\begin{align} 
A &:= \frac{ 2\kappa \omega^2 \Omega }{ 
  \bigl[ (\omega-2\Omega)^2 + \kappa^2 \bigr] 
  \bigl[ (\omega+2\Omega)^2 + \kappa^2 \bigr] }, \\ 
B &:= \frac{ \frac{1}{2} \omega^2 \bigl( \omega^2 - 4\Omega^2 +
  \kappa^2 \bigr) }{ 
  \bigl[ (\omega-2\Omega)^2 + \kappa^2 \bigr] 
  \bigl[ (\omega+2\Omega)^2 + \kappa^2 \bigr] }, \\ 
C &:= \frac{\omega^2}{6(\omega^2 + \kappa^2)}. 
\end{align} 
\end{subequations} 
These expressions reveal that a resonance occurs when 
$2\Omega = \omega$; the factor of two is associated with the
quadrupolar nature of the tidal interaction. The width of the 
resonant features is determined by $\kappa$, the inverse of the
damping time. Equation (\ref{F_expression}) can be inserted into 
Eq.~(\ref{tidalresp1}) to obtain the body's quadrupole-moment tensor. 
(Strictly speaking, the integration from 
$t' = -\infty$ to $t' = t$ in Eq.~(\ref{tidalresp2}) does not converge, 
because the particle is assumed to be at all times on a fixed circular 
orbit. This defect can be remedied by choosing the lower bound to be 
some finite reference time $t_0$. The resulting dependence of 
${\cal F}_{ab}$ on $t_0$ can then be shown to correspond to a transient 
response that decays exponentially. The expression shown above 
corresponds to the steady-state response that emerges when 
$t \gg t_0$.)  

\subsection{Self-force} 

\begin{figure}
\subfloat{\includegraphics[width=0.48\textwidth]{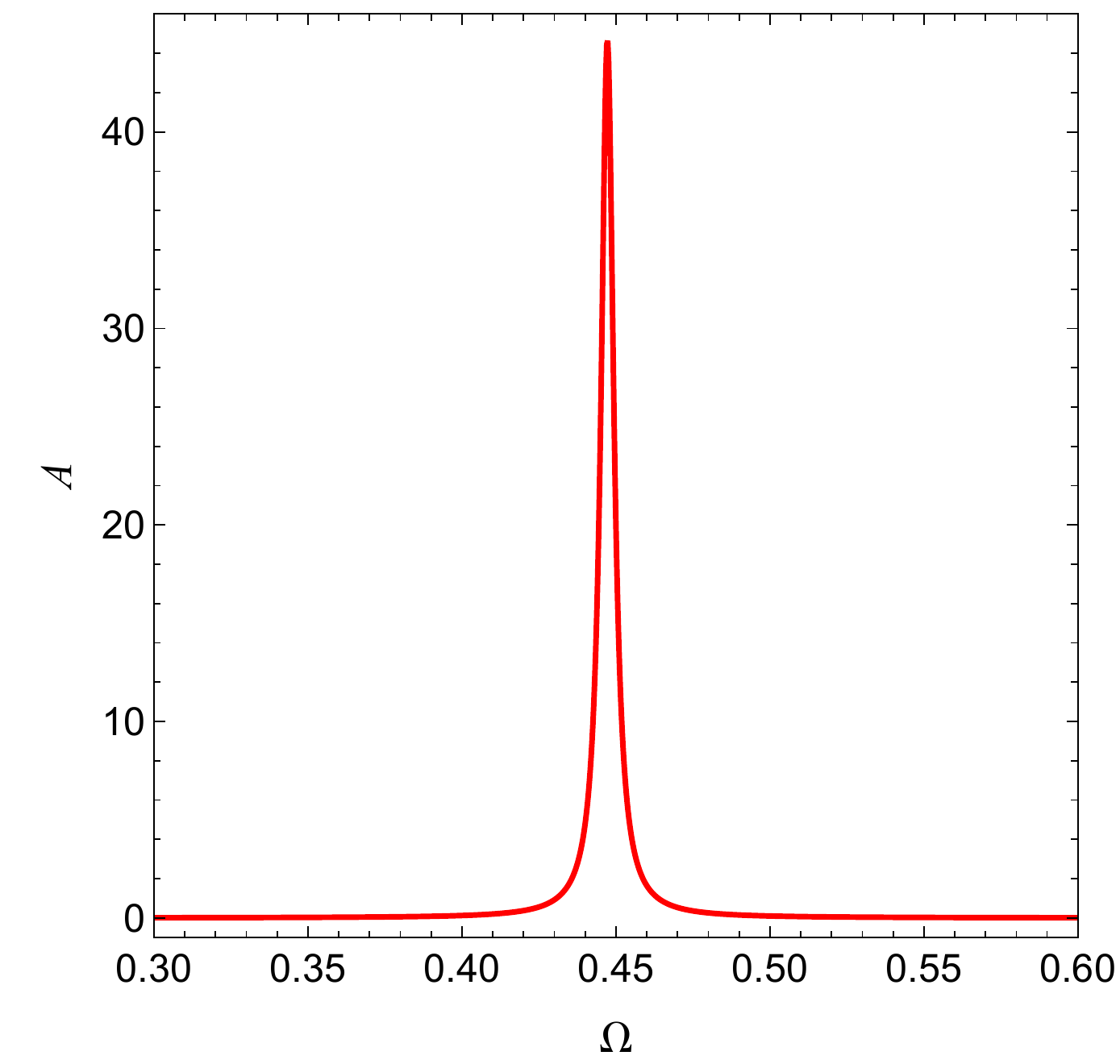}}
    \hfill
\subfloat{\includegraphics[width=0.48\textwidth]{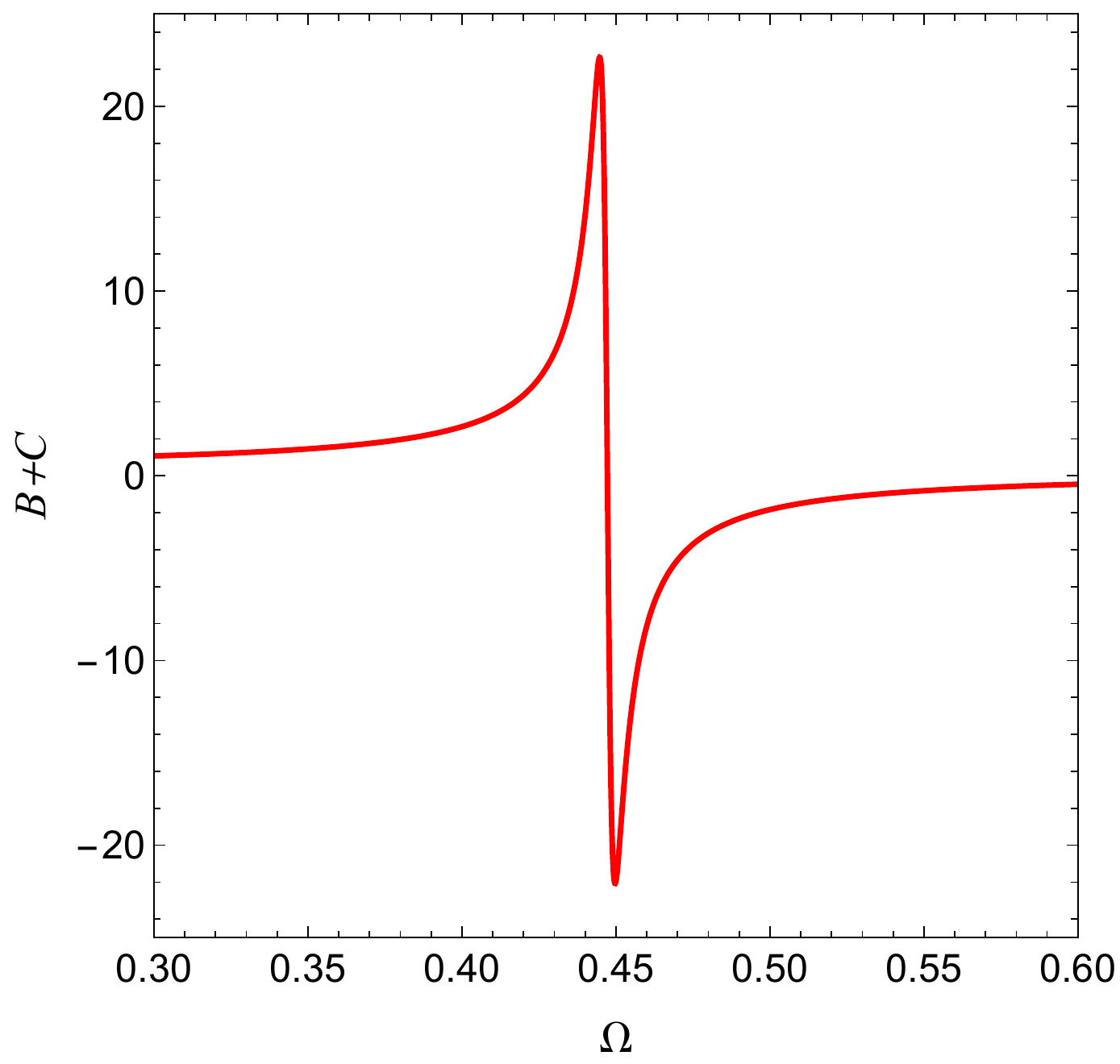}}
\caption{Plots of $A$ and $B+C$ as a function of $\Omega$. The orbital
  frequency is presented in units of $\sqrt{M/R^3}$, and the resonance
  occurs when $\Omega = \frac{1}{2}\omega \simeq 0.447$. The damping
  parameter is set to $\kappa = 0.005$, in the same frequency units.}
\label{fig:fig5}
\end{figure}
 
The tide produces a deformation of the body's gravitational potential,
which is otherwise given by $U = M/r$. According to Eq.~(1.149) of
Ref.~\cite{poisson-will:14}, the perturbation is described by 
\begin{equation} 
\delta U = \frac{1}{2} I^{ab} \partial_{ab} 
\biggl( \frac{1}{r} \biggr) 
\end{equation} 
when it is truncated to the leading, quadrupole order. This
perturbation contributes a force $\f_a = \m \partial_a \delta U$
acting on the particle, and this shall be our Newtonian model for the
self-force. The self-force is to be added to the original force 
$\m \partial_a U$ created by the body, which is responsible for 
maintaining the circular orbit. A straightforward computation, 
making use of Eq.~(\ref{tidalresp1}) as well as Eq.~(1.152c) of
Ref.~\cite{poisson-will:14}, returns 
\begin{equation} 
\f_a = -\frac{1}{5} \m M R^2 {\cal F}^{bc} 
\partial_{abc} \biggl( \frac{1}{r} \biggr) 
= 3 \m M \frac{R^2}{r^4} {\cal F}^{bc} n_\stf{abc}, 
\end{equation} 
where $n_\stf{abc} := n_a n_b n_c - \frac{1}{5}( n_a \delta_{bc} 
+ n_b \delta_{ac} + n_c \delta_{ab})$. Inserting
Eq.~(\ref{F_expression}) and performing the tensorial manipulations
yields 
\begin{equation} 
\bm{\f} = -\frac{9}{4} \frac{\m^2 R^5}{r^7} \Bigl[ 
3 (B+C) \bm{n} + 2 A \bm{\lambda} \Bigr]. 
\label{Newt_sf} 
\end{equation} 
The self-force features a radial component proportional to $B+C$, and
a tangential component proportional to $A$; these functions of the
orbital frequency $\Omega$ are plotted in Fig.~\ref{fig:fig5}. 
The radial component of the self-force is
associated with conservative effects, while its tangential component
is associated with dissipative effects. The components of the
Newtonian self-force, $\f_r \propto -(B+C)$ and 
$\f_\phi \propto -A$, can be compared across a resonance to those of
the relativistic self-force in Fig.~\ref{fig:fig2}. We have
qualitative agreement, and in fact, we have verified that the
Newtonian expressions provide excellent fits of each resonant feature 
when $\omega$ and $\kappa$ are matched to the relevant quasinormal
mode.   

\subsection{Orbital evolution} 

We next wish to describe the impact of the self-force on the motion of
the particle. The equations of motion are $\bm{F} = \m \bm{a}$, where  
$\bm{F}$ is the total force acting on the particle, and 
\begin{equation} 
\bm{a} = \bigl( \ddot{r} - r \Omega^2 \bigr) \bm{n}
+ \frac{1}{r} \frac{d}{dt} \bigl( r^2 \Omega \bigr) \bm{\lambda} 
\end{equation} 
is the acceleration vector; $\Omega := \dot{\phi}$, and an overdot
indicates differentiation with respect to $t$. To formalize the fact
that the self-force is small compared with the force produced by the
unperturbed body, we introduce a small parameter $\epsilon \ll 1$ and
expand the total force in powers of $\epsilon$,  
\begin{equation} 
\bm{F} = \bm{F}_0 + \epsilon \bm{F}_1 + O(\epsilon^2).
\end{equation} 
Here $\bm{F}_0 = -\m M \bm{n}/r^2$ and 
$\epsilon \bm{F}_1 = \bm{\f}$. We assume that the particle follows
a circular orbit that changes very slowly as a result of the
self-force. To formalize this assumption we follow Pound's multiscale 
methods (see Sec.~IV of Ref.~\cite{pound:15}) and introduce a slow-time
variable $\bar{t} := \epsilon t$, and state that the orbital radius
$r$ and the angular frequency $\Omega$ shall be functions of
$\bar{t}$; quantities (such as the velocity vector) that vary over the
orbital time scale are written as functions of the fast-time variable
$t$. We assume also that the orbital frequency admits an expansion in 
powers of $\epsilon$: 
\begin{equation} 
\Omega = \Omega_0(\bar{t}) + \epsilon \Omega_1(\bar{t}) 
+ O(\epsilon^2). 
\end{equation} 

To obtain a description of the perturbed motion we insert these 
functional relations into the equations of motion (writing, for
example, $\dot{r} = \epsilon r'$, with a prime indicating
differentiation with respect to $\bar{t}$), and we expand in powers of
$\epsilon$. The radial component becomes  
\begin{equation} 
\m \bigl( r \Omega_0^2 + 2\epsilon r \Omega_0 \Omega_1 \bigr) 
= \m M/r^2 - \epsilon \bm{F}_1 \cdot \bm{n} 
+ O(\epsilon^2), 
\end{equation} 
while the tangential component  becomes 
\begin{equation} 
\epsilon \m \bigl( r\Omega_0' + 2\Omega_0 r' \bigr) 
= \epsilon \bm{F}_1 \cdot \bm{\lambda}
+ O(\epsilon^2). 
\end{equation} 
At order $\epsilon^0$ we recover the relation $\Omega_0^2 = M/r^3$,
which implies that $r'/r = -(2/3) \Omega_0'/\Omega_0$. At order
$\epsilon$ the radial equation produces 
\begin{equation} 
\Omega_1 = -\frac{1}{2\m r \Omega_0} \bm{F}_1 \cdot \bm{n}, 
\end{equation} 
while the tangential equation gives rise to
\begin{equation} 
\Omega_0' = -\frac{3}{\m r} \bm{F}_1 \cdot \bm{\lambda}. 
\end{equation}
These are the required evolution equations. To display them in their
final form we eliminate $\bar{t}$ in favor of $t$ and re-express
$\bm{F}_1$ in terms of the self-force $\bm{\f}$. We obtain the
complete set of equations 
\begin{subequations} 
\begin{align} 
\Omega &= \Omega_0 - \frac{1}{2\m r \Omega_0} \bm{\f} \cdot \bm{n}, \\
 \frac{d\Omega_0}{dt} &= -\frac{3}{\m r} \bm{\f} \cdot \bm{\lambda}, \\ 
\Omega_0^2 &= \frac{M}{r^3}.
\end{align} 
\end{subequations} 
One sees that in this approximate description of the motion, the
radial component of the self-force produces a shift in angular
frequency with respect to $\Omega_0$, and that the tangential
component drives an evolution of $\Omega_0$ (and therefore of $r$).  

\begin{figure} 
\includegraphics[width=0.8\linewidth]{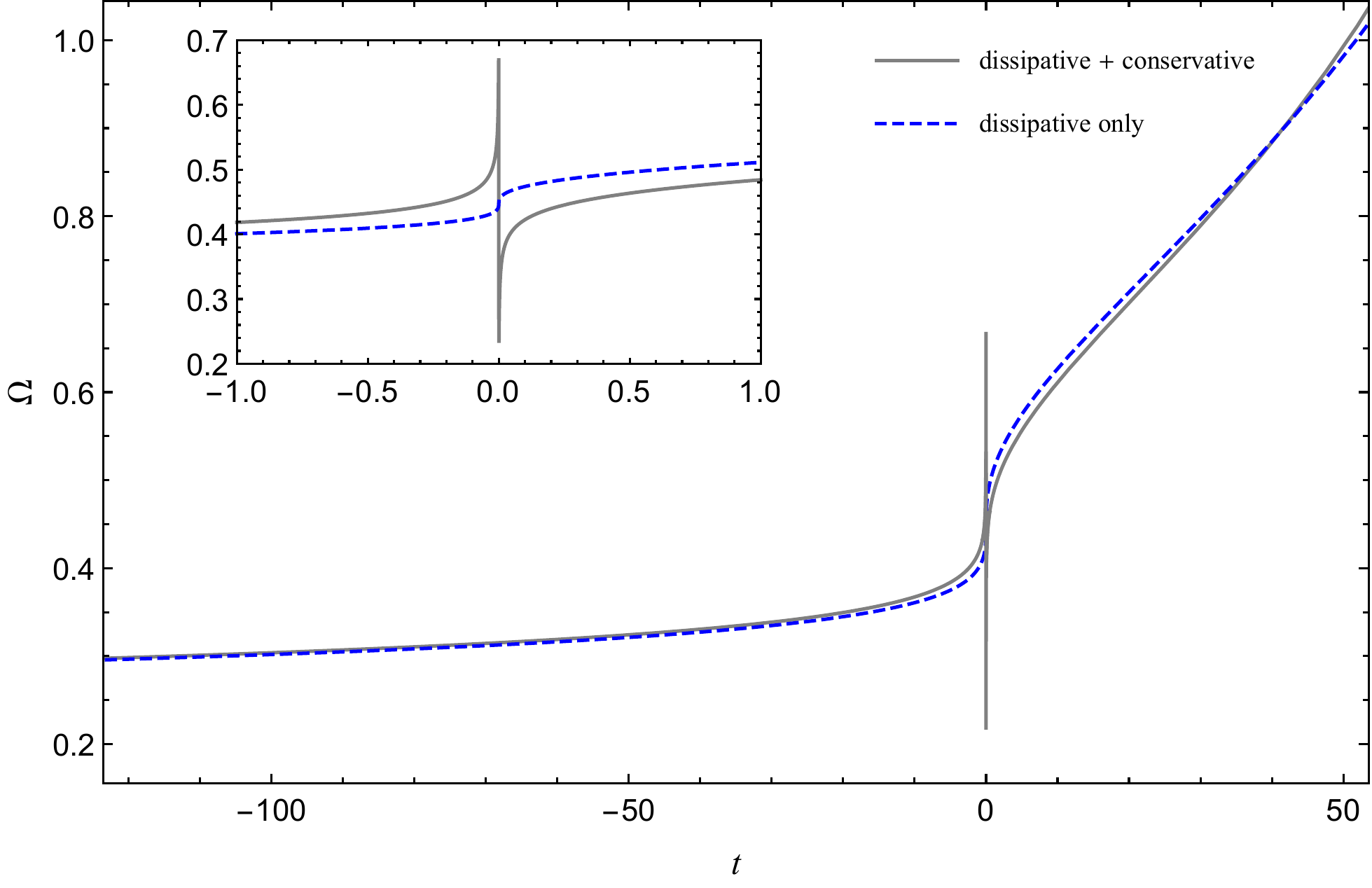}
\caption{Plot of the angular velocity $\Omega$ as a function of time.   
The frequency is presented in units of $\sqrt{M/R^3}$, and the resonance
  occurs when $\Omega = \frac{1}{2}\omega \simeq 0.447$. The damping 
  parameter is set to $\kappa = 0.005$, in the same frequency units. Time is
  presented in units of $(M/\m) \sqrt{R^3/M}$, and the zero of time
  coincides with resonance crossing. The mass ratio was set to $\m/M =
  0.1$ to ensure that the curves can be distinguished.}  
\label{fig:fig6} 
\end{figure} 
 
In our current application the self-force is given by
Eq.~(\ref{Newt_sf}), and the evolution equations become 
\begin{subequations} 
\begin{align} 
\Omega &= \Omega_0 + \frac{27}{8} \frac{\m R^5}{\Omega_0 r^8} (B+C),
\\ 
\frac{d\Omega_0}{dt} &= \frac{27}{2} \frac{\m R^5}{r^8} A. 
\end{align} 
\end{subequations} 
Before integrating these equations it is convenient to rescale the
variables so as to make them dimensionless. We shall therefore measure
all frequencies in units of $\Omega^* := \sqrt{M/R^3}$, measure time
in units of $t^* :=(M/\m) \sqrt{R^3/M}$, and measure $r$ in units of
$R$. In these natural units integration yields 
\begin{subequations} 
\begin{align} 
t &= t_i + \frac{2}{27} \frac{1}{\kappa \omega^2} \biggl\{ 
\frac{1}{\Omega_i^{16/3}} \biggl[ \frac{3}{32} (\omega^2 + \kappa^2)^2 
- \frac{6}{5} (\omega^2-\kappa^2) \Omega_i^2 + 6\Omega_i^4 \biggr] 
\nonumber \\ & \quad \mbox{} 
-  \frac{1}{\Omega_0^{16/3}} \biggl[ \frac{3}{32} (\omega^2 + \kappa^2)^2 
- \frac{6}{5} (\omega^2-\kappa^2) \Omega_0^2 + 6\Omega_0^4 \biggr]
\biggr\}, \\ 
\Omega &= \Omega_0 + \frac{27}{8} \frac{\m}{M} \Omega_0^{13/3} (B+C), 
\end{align} 
\end{subequations} 
where $t_i$ is the time at which $\Omega_0 = \Omega_i$. These
equations provide a parametric expression for $\Omega(t)$, with
$\Omega_0$ playing the role of parameter. 

A plot of the angular frequency is presented in
Fig.~\ref{fig:fig6}. The dashed blue curve represents $\Omega_0(t)$,
which changes with time by purely dissipative effects; the change is
slow off resonance, and rapid near resonance. The solid gray curve
represents $\Omega(t)$, which also includes the shift produced by the
conservative piece of the self-force. The shift is positive before resonance, and negative just after
it. The inset shows the details of resonance crossing, with
the rapid increase of $\Omega(t)$ immediately prior to resonance, the
very sudden decrease at resonance, and the rapid final increase
immediately after resonance. Figure~\ref{fig:fig6} can be compared
with its relativistic version in Fig.~\ref{fig:fig3}; we again find
excellent qualitative agreement. 

The radial (conservative) component of the self-force produces a shift
$\Delta \Omega := \Omega - \Omega_0$ of the angular frequency relative
to $\Omega_0$, which evolves by sole virtue of the tangential
(dissipative) component of the self-force. This shift produces a
dephasing $\Delta \phi$ of the orbit relative to the purely
dissipative evolution. The dephasing per frequency interval is given by   
\begin{equation} 
\frac{d \Delta \phi}{d\Omega_0} 
= \frac{d \Delta \phi/dt}{d\Omega_0/dt} 
= \frac{\Delta\Omega}{d\Omega_0/dt}, 
\end{equation} 
and this evaluates to 
\begin{equation} 
\frac{d \Delta \phi}{d\Omega_0} = \frac{B+C}{4\Omega_0 A} 
= \frac{\omega^2+\kappa^2}{12\kappa\Omega_0^2} 
- \frac{5\omega^2+\kappa^2}{12\kappa(\omega^2+\kappa^2)} 
+ \frac{\Omega_0^2}{3\kappa(\omega^2+\kappa^2)}. 
\end{equation} 
The total dephasing across the resonance can be obtained by
integrating this expression over the frequency interval $\frac{1}{2}
\omega -\kappa < \Omega_0 < \frac{1}{2}\omega + \kappa$, with 
$\kappa$ providing a measure of the width of the resonance. A simple
calculation shows that the total dephasing amounts to $\frac{73}{18}
(\kappa/\omega)^2$ when $\kappa/\omega \ll 1$. This is small, which 
was to be expected from the fact that the large increase in angular
velocity witnessed immediately before and after resonance is almost
completely cancelled by the rapid decrease during resonance.  
  
\section{Scalar gravity coupled to a perfect fluid} 
\label{sec:equations} 

In this section we introduce the scalar theory of gravity adopted in
this work. The theory is inspired from scalar-tensor theories of
gravity, in which we simply make the tensor component of the
gravitational field nondynamical. The theory, in fact, is identical 
to the venerable Nordstr\"om theory, which is conveniently reviewed 
in Ref.~\cite{deruelle:11}. 

The field equation for the gravitational field $\Phi$ and the equations of motion for the matter fields are initially formulated
in a curved spacetime with a nondynamical metric $g_{\alpha\beta}$, and they are postulated to be of the form
\begin{equation} 
\Box \Phi = 4\pi \alpha(\Phi) T^\mu_{\ \mu}, \qquad 
\nabla_\beta T^{\alpha\beta} 
= -\alpha(\Phi) T^\mu_{\ \mu} \nabla^\alpha \Phi, 
\label{eqns} 
\end{equation} 
in which $\Box := g^{\alpha\beta} \nabla_\alpha \nabla_\beta$ is the 
covariant wave operator, $\alpha(\Phi)$ is an arbitrary function of
the scalar potential, and $T^{\alpha\beta}$ is the energy-momentum
tensor of the matter fields. 

The field and matter equations (\ref{eqns}) can be derived on the 
basis of a variational principle. Consider the action
\begin{equation}
S = \frac{1}{16 \pi} \int d^4 x \sqrt{-g} \left( \mathrm{R} - 2 g^{\alpha \beta} \nabla_\alpha \Phi \nabla_\beta \Phi \right) + S_m[\Psi_m, \beta(\Phi) g_{\alpha\beta}],
\end{equation}
where $g:=\det(g_{\alpha \beta})$, $\mathrm{R}$ is the Ricci scalar,
and $S_m$ is the action of the matter fields, which are
collectively denoted by $\Psi_m$ and couple to the conformally
rescaled metric $\tilde{g}_{\alpha\beta}:= \beta(\Phi)g_{\alpha\beta}$,
with $\beta(\Phi)$ an arbitrary function of $\Phi$.
Variation with respect to $g_{\alpha\beta}$ yields
\begin{equation}
G^{\alpha \beta} - 2 \nabla^\alpha \Phi \nabla^\beta \Phi +
g^{\alpha \beta} \nabla_\mu \Phi \nabla^\mu \Phi = 8\pi T^{\alpha \beta},
\label{Gmunu}
\end{equation}
where $T^{\alpha \beta} := (2/\sqrt{-g}) \delta S_m [\Phi_m, \beta(\Phi) g_{\mu\nu}]/\delta g_{\alpha \beta} $. 
Variation with respect to the scalar field gives
\begin{equation}
\Box \Phi = - 2 \pi \frac{d\ln\beta(\Phi)}{d\Phi} T^\mu_{\ \mu},
\label{eqPhi}
\end{equation}
which is the field equation displayed in Eq.~(\ref{eqns}) upon 
identifying $\alpha = -(1/2) d\ln\beta/d\Phi$. Taking the 
divergence of Eq.~(\ref{Gmunu}), and making use of 
Eq.~(\ref{eqPhi}), one obtains the equations of motion for the
matter fields shown in Eq.~(\ref{eqns}). Notice, from Eq.~(\ref{Gmunu}),
that the field equations admit a conserved total energy-momentum 
tensor, the sum of $T^{\alpha\beta}$ and 
\begin{equation} 
t^{\alpha\beta} := \frac{1}{4\pi} \Bigl( \nabla^\alpha \Phi
\nabla^\beta \Phi - \frac{1}{2} g^{\alpha\beta} \nabla_\mu \Phi 
\nabla^\mu \Phi \Bigr), 
\end{equation} 
which can be interpreted as the energy-momentum tensor of the
gravitational field. 
The construction above parallels the Einstein-frame formulation of
scalar-tensor theories \cite{damour-espositofarese:92}, 
which are viable theories of 
gravity for some range of coupling functions. On the other hand,
in our simplified model we take Eqs.~(\ref{eqns}) as the starting
point, assuming a fixed nondynamical background metric.

We now consider a perfect fluid with rest-mass density $\rho$, 
pressure $p$, density of internal energy $\epsilon$, total energy
density $\mu = \rho + \epsilon$, and velocity $u^\alpha$ coupled to the scalar gravitational field $\Phi$. With 
\begin{equation} 
T^{\alpha\beta} = \mu u^\alpha u^\beta + p P^{\alpha\beta}, \qquad 
P^{\alpha\beta} = g^{\alpha\beta} + u^\alpha u^\beta,
\end{equation} 
the field and fluid equations (\ref{eqns}) take the explicit form 
\begin{subequations} 
\begin{align}
0 &= S := \Box \Phi + 4\pi \alpha (\mu - 3p), 
\label{wave} \\ 
0 &= S' := u^\alpha \nabla_\alpha \mu + (\mu + p) \nabla_\alpha u^\alpha 
+ \alpha (\mu - 3p) u^\alpha \nabla_\alpha \Phi, 
\label{firstlaw1} \\ 
0 &= S_\alpha := (\mu+p) a_\alpha + P_\alpha^{\ \beta} \nabla_\beta p 
- \alpha (\mu-3p) P_\alpha^{\ \beta} \nabla_\beta \Phi, 
\label{euler} 
\end{align} 
\end{subequations} 
where $a_\alpha := u^\beta \nabla_\beta u_\alpha$ is the fluid's
covariant acceleration. To these equations we adjoin an equation of
state $p = p(\rho, s)$, $\epsilon = \epsilon(\rho,s)$, in which 
$s$ is the fluid's specific entropy (entropy per unit mass), and the 
statement of rest-mass conservation,  
\begin{equation} 
\nabla_\alpha \bigl( \rho u^\alpha \bigr) = 0.  
\label{mass_cons} 
\end{equation} 
Equation (\ref{firstlaw1}) can then be expressed in the form of a 
dynamical version of the first law of thermodynamics,  
\begin{equation} 
\rho T\, \frac{ds}{d\tau} 
= \frac{d\epsilon}{d\tau} 
- \frac{\epsilon+p}{\rho}\, \frac{d\rho}{d\tau} 
= \frac{d\mu}{d\tau} 
- \frac{\mu+p}{\rho}\, \frac{d\rho}{d\tau} 
= -\alpha (\mu - 3p) \frac{d\Phi}{d\tau}, 
\label{firstlaw2} 
\end{equation} 
where $T$ is the fluid's temperature and $\tau$ is proper time on
the world line of each fluid element. We see that the gravitational field
is a source of heat when the fluid configuration depends on time.   

In the sequel we shall simplify the formulation of the theory by
setting 
\begin{equation} 
\alpha(\Phi) = 1, \qquad 
g_{\alpha\beta} = \eta_{\alpha\beta},  
\label{simpli} 
\end{equation} 
in which $\eta_{\alpha\beta}$ is the Minkowski metric in the selected 
coordinate system (which may not be the standard Lorentzian
coordinates). In this formulation the theory of gravity is linear, and
the background geometry is flat. 

\section{Stellar models} 
\label{sec:star} 

With the choice made in Eq.~(\ref{simpli}) we consider static and
spherically-symmetric configurations of the fluid and gravitational
field. Adopting spherical polar coordinates $(t,r,\theta,\phi)$ and
letting $u^\alpha = (1,0,0,0)$, $a_\alpha = 0$, we find that the
field and fluid equations reduce to the system 
\begin{subequations} 
\label{static_SS} 
\begin{align} 
\frac{dm}{dr} &= 4\pi r^2 (\mu - 3p), \\ 
\frac{d\Phi}{dr} &= -\frac{m(r)}{r^2}, \\ 
\frac{dp}{dr} &= (\mu-3p) \frac{d\Phi}{dr} = -(\mu-3p) \frac{m(r)}{r^2}, 
\end{align} 
\end{subequations} 
in which an effective mass function $m(r)$ was introduced as an
auxiliary variable. The structure equations closely resemble the
Newtonian equations, with the effective mass density 
$\mu - 3p = \rho + \epsilon - 3p$ replacing the rest-mass density
$\rho$. The equations guarantee that $m(r) = M = \mbox{constant}$
outside the body, where the potential is given by $\Phi = M/r$.  

A particularly simple stellar structure follows if we postulate the  
relation $p = K (\rho + \epsilon - 3 p)^2$ between the pressure and
the effective mass density, where $K$ is a constant; this is a slight
modification of the $n=1$ polytropic model described by 
$p = K \rho^2$. This modified assignment gives rise to a physically
acceptable, zero-temperature equation of state, which we now write as
$\rho = \rho(p)$ and $\epsilon = \epsilon(p)$. 

The expression for $\epsilon$ can be found by integrating 
$d\epsilon = (\epsilon+p)\rho^{-1}\, d\rho$ with 
$\rho = \sqrt{p/K} - \epsilon + 3p$. This yields  
\begin{subequations} 
\label{n1_EOS} 
\begin{align}    
\rho &= \sqrt{\frac{p}{K}} \sqrt{1 + 4x}, \\ 
\epsilon &= 3p + \sqrt{\frac{p}{K}} \Bigl( 1 - \sqrt{1 + 4x} \Bigr),
\end{align}
\end{subequations} 
with $x := \sqrt{Kp}$. The speed of sound associated with this
equation of state is given by 
\begin{equation} 
c_s^2 = \frac{dp}{d\mu} = \frac{2x}{1+6x} < \frac{1}{3}.  
\end{equation} 
When $x$ is small the equation of state reduces to 
$\rho = \sqrt{p/K}[1 + 2x + O(x^2)]$ and 
$\epsilon = p[1 + 2x + O(x^2)]$, and it deviates little from a 
polytropic equation of state with $n=1$.  

The solution to the structure equations (\ref{static_SS}) 
for this equation of state can be expressed as  
\begin{subequations} 
\label{structure_n1} 
\begin{align} 
p &= \frac{\pi M^2}{8 R^4} \biggl[ \frac{ \sin(\pi r/R) }{ \pi r/R }
\biggr]^2, \\ 
\rho &= \frac{\pi M}{4 R^3} \frac{ \sin(\pi r/R) }{ \pi r/R } 
\sqrt{ 1 + \frac{2M}{R} \frac{ \sin(\pi r/R) }{ \pi r/R } }, \\ 
\mu &= \frac{\pi M}{4 R^3}\frac{ \sin(\pi r/R) }{ \pi r/R }
\biggl[ 1 + \frac{3M}{2R} \frac{ \sin(\pi r/R) }{ \pi r/R } \biggr], \\ 
m &= M \frac{r}{R} \biggl[ \frac{ \sin(\pi r/R) }{ \pi r/R } 
- \cos(\pi r/R) \biggr]; 
\end{align} 
\end{subequations} 
the constant $K$ is related to the stellar radius $R$ by $K = (2/\pi)
R^2$. The pressure is largest at the center of the star, where it is
equal to $p_c = \pi M^2/(8 R^4)$. For future reference we record that  
\begin{equation} 
\Gamma := \frac{\rho}{p} \frac{dp/dr}{d\rho/dr} 
= 2 \frac{\pi r + 2M \sin(\pi r/R)}{\pi r + 3M \sin(\pi r/R)}.  
\label{Gamma_n1} 
\end{equation} 
This function increases monotonically with $r$, and it is bounded by 
\begin{equation} 
\Gamma(r=0) = 2 \frac{1+2M/R}{1+3M/R} \leq \Gamma(r) 
\leq 2 = \Gamma(r=R). 
\end{equation} 
 
\section{Fluid perturbations} 
\label{sec:perturbed} 

In this section we consider perturbations of the equilibrium 
stellar model of Sec.~\ref{sec:star} and describe the quasinormal
modes of the fluid-gravity system.

\subsection{Formalism} 
\label{subsec:perturbed1} 

The unperturbed state of the fluid is taken to be an equilibrium for
which $u^\alpha = (1,0,0,0)$, $a_\alpha = 0$, and all fluid variables
are independent of time. The perturbation, however, is allowed to
depend on time. We continue to work in Minkowski spacetime, but keep 
the equations covariant in order to retain the freedom to choose the
coordinate system. We rely on the Lagrangian theory of fluid
perturbations reviewed in Ref.~\cite{friedman-stergioulas:13}, and
import many relevant results from this reference. We recall that for
any fluid variable $Q$, scalar or tensorial, the Lagrangian
perturbation $\Delta Q$ and Eulerian perturbation $\delta Q$ are
related by $\Delta Q = \delta Q + {\cal L}_\xi Q$, in which 
${\cal L}_\xi$ is the Lie derivative in the direction of the
Lagrangian displacement vector $\xi^\alpha$.   

As derived in Sec.~2.2 of Ref.~\cite{friedman-stergioulas:13}, the
Lagrangian perturbation of the velocity vector is given by 
\begin{equation} 
\Delta u^\alpha = \frac{1}{2} u^\alpha u^\beta u^\gamma 
\Delta \eta_{\beta\gamma}, 
\label{Delta_u} 
\end{equation} 
while the perturbation of the rest-mass density is 
\begin{equation} 
\Delta \rho = -\frac{1}{2} \rho P^{\alpha\beta} 
\Delta \eta_{\alpha\beta}.  
\label{Delta_rho} 
\end{equation} 
In our case there is no Eulerian perturbation of the metric, and
$\Delta \eta_{\alpha\beta} = \nabla_\alpha \xi_\beta 
+ \nabla_\beta \xi_\alpha$. 

To derive an expression for $\Delta \mu$ we rely on the dynamical 
formulation of the first law of thermodynamics, Eq.~(\ref{firstlaw2}),
which becomes 
\begin{equation} 
0 = \Delta \frac{d\mu}{d\tau} 
- \frac{\mu+p}{\rho}\, \Delta \frac{d\rho}{d\tau} 
+ (\mu - 3p) \Delta \frac{d\Phi}{d\tau}
\end{equation} 
after the perturbation; the terms proportional to $d\rho/d\tau$ and
$d\Phi/d\tau$ were eliminated because these quantities vanish in the
unperturbed state. The commutation relation 
\begin{equation} 
\biggl[ \Delta, \frac{d}{d\tau} \biggr] f = \frac{1}{2} \frac{df}{d\tau} 
u^\alpha u^\beta\, \Delta \eta_{\alpha\beta},  
\end{equation} 
where $f$ is an arbitrary scalar, further implies that the first law can be
written as 
\begin{equation}  
0 = \frac{d}{d\tau} \Delta \mu  
- \frac{\mu+p}{\rho}\, \frac{d}{d\tau} \Delta \rho 
+ (\mu - 3p) \frac{d}{d\tau} \Delta \Phi
= \frac{d}{d\tau} \biggl[ \Delta \mu  
- \frac{\mu+p}{\rho}\, \Delta \rho 
+ (\mu - 3p) \Delta \Phi \biggr].  
\end{equation} 
Integrating, we arrive at 
\begin{equation} 
\Delta \mu = \frac{\mu+p}{\rho}\, \Delta \rho 
- (\mu - 3p) \Delta \Phi
= -\frac{1}{2} (\mu+p) P^{\alpha\beta}\, \Delta \eta_{\alpha\beta} 
- (\mu-3p) \Delta \Phi, 
\label{Delta_mu} 
\end{equation} 
where $\Delta \Phi = \delta \Phi + \xi^\alpha \nabla_\alpha \Phi$. 

Turning next to $\Delta p$, we invoke the equation of state 
$p = p(\rho,s)$ to write 
\begin{align} 
\frac{dp}{d\tau} &= 
\biggl(\frac{\partial p}{\partial \rho} \biggr)_s\, \frac{d\rho}{d\tau}
+ \biggl(\frac{\partial p}{\partial s} \biggr)_\rho\, \frac{ds}{d\tau} 
\nonumber \\ 
&= \frac{\Gamma_1 p}{\rho}\, \frac{d\rho}{d\tau} 
+ (\Gamma_3-1) \rho T\, \frac{ds}{d\tau} 
\nonumber \\ 
&= \frac{\Gamma_1 p}{\rho}\, \frac{d\rho}{d\tau} 
- (\Gamma_3-1) (\mu-3p)\, \frac{d\Phi}{d\tau},  
\end{align} 
where 
\begin{equation} 
\Gamma_1 := \biggl( \frac{\partial \ln p}{\partial \ln \rho}
\biggr)_s, \qquad    
\Gamma_3 - 1 := \biggl( \frac{\partial \ln T}
{\partial \ln \rho} \biggr)_s 
\label{adiab_exp} 
\end{equation} 
are the standard adiabatic exponents. To go from the first to the
second line we invoked the Maxwell relation 
$(\partial p/\partial s)_\rho 
= \rho^2(\partial T/\partial \rho)_s$, and to go to the third
line we used Eq.~(\ref{firstlaw2}) to relate $ds/d\tau$ to
$d\Phi/d\tau$. Perturbing and integrating the equation for $dp/d\tau$, 
as we did for $d\mu/d\tau$, we arrive at 
\begin{equation} 
\Delta p = \frac{\Gamma_1 p}{\rho}\, \Delta \rho 
- (\Gamma_3-1) (\mu - 3p) \Delta \Phi
= -\frac{1}{2} \Gamma_1 p P^{\alpha\beta}\, \Delta \eta_{\alpha\beta} 
- (\Gamma_3-1) (\mu-3p) \Delta \Phi.  
\label{Delta_p} 
\end{equation} 

Importing from Sec.~7.1 of Ref.~\cite{friedman-stergioulas:13} 
and keeping in mind that the unperturbed state is an equilibrium
configuration, the Lagrangian perturbation of the acceleration vector
is  
\begin{equation} 
\Delta a_\alpha = P_\alpha^{\ \beta} \biggl[ 
u^\mu {\cal L}_u \Delta \eta_{\beta\mu} 
- \frac{1}{2} \nabla_\beta \bigl( u^\mu u^\nu\, 
\Delta \eta_{\mu\nu} \bigr) \biggr].  
\label{Delta_a} 
\end{equation} 
We also have 
\begin{equation} 
\Delta P_\alpha^{\ \beta} = u^\beta P_\alpha^{\ \gamma} u^\delta 
\Delta \eta_{\gamma\delta}  
\label{Delta_P} 
\end{equation} 
and 
\begin{equation} 
\Delta \nabla_\beta p = \nabla_\beta \Delta p, \qquad 
\Delta \nabla_\beta \Phi = \nabla_\beta \Delta \Phi.  
\label{Delta_gradp} 
\end{equation} 
With all this we arrive at the perturbation of Euler's equation
(\ref{euler}),
\begin{equation} 
0 = \Delta S_\alpha = P_\alpha^{\ \beta} \biggl[ 
(\mu+p) {\cal L}_u \bigl( u^\mu\, \Delta \eta_{\beta\mu} \bigr) 
- \frac{1}{2}(\mu+p) \nabla_\beta \bigl( u^\mu u^\nu\, 
  \Delta \eta_{\mu\nu} \bigr) 
+ \nabla_\beta \Delta p 
- (\mu - 3p) \nabla_\beta \Delta \Phi 
- (\Delta \mu - 3 \Delta p) \nabla_\beta \Phi \biggr]. 
\label{Delta_euler} 
\end{equation} 
The (Eulerian) perturbation of the gravitational field equation 
gives 
\begin{equation} 
0 = \delta S := \Box \delta \Phi + 4\pi (\delta\mu - 3\delta p),  
\label{delta_wave} 
\end{equation} 
and these equations govern the behavior of the perturbed
configuration, as described by $\xi^\alpha$ and $\delta \Phi$.  

\subsection{Implementation} 
\label{subsec:perturbed2}

We take the unperturbed configuration to be static and 
spherically-symmetric, so that the body's structure is determined by 
Eqs.~(\ref{static_SS}). The components of the Lagrangian 
displacement vector can be decomposed in a sum over modes, each of
them of the form
\begin{equation} 
\xi_t = 0, \qquad 
\xi_r = x(r) Y_{\ell m}(\theta,\phi) e^{-i\omega t}, \qquad 
\xi_A = z(r) \partial_A Y_{\ell m}(\theta,\phi) e^{-i\omega t}, 
\label{xi_components} 
\end{equation} 
where the uppercase index $A = 1, 2$ runs over the angular
coordinates $\theta^A = (\theta,\phi)$, and $Y_{\ell m}(\theta,\phi)$
are the usual spherical-harmonic functions; we have that $z$ is 
undefined when $\ell = 0$. The Eulerian perturbation of the 
gravitational potential is similarly expressed as  
\begin{equation} 
\delta \Phi = P(r) Y_{\ell m}(\theta,\phi) e^{-i\omega t}. 
\end{equation} 
In these equations, it is understood (but not shown) that the
variables $\{ x, z, P \}$ carry a multipole label $\ell$ and a
frequency label $\omega$; $m$ is not required as a label because the
unperturbed configuration is spherically-symmetric.   

Substitution of these expressions into the perturbation equations 
$\Delta S_\alpha = 0$ and $\delta S = 0$ produces an explicit system
of ordinary differential equations for the variables $\{ x, z, P\}$. 
However, it turns out to be more convenient to formulate the 
perturbation equations in terms of a new variable $y(r)$ defined by    
\begin{equation} 
\delta p = y(r) Y_{\ell m}(\theta,\phi) e^{-i\omega t}. 
\end{equation} 
As a consequence of Eq.~(\ref{Delta_p}) and the angular components of
the perturbed Euler equation, we find that $y$ is related to $z$ by 
\begin{equation} 
y = \omega^2 (\mu+p) z + (\mu-3p) P. 
\label{y_vs_z} 
\end{equation} 
Assuming that $\omega \neq 0$ (this special case will be considered
below), the resulting equations can be manipulated to take the
schematic form
\begin{subequations} 
\label{xyP_system} 
\begin{align} 
0 &= dx/dr + (\mbox{\ }) x + (\mbox{\ }) y + (\mbox{\ }) P, \\
0 &= dy/dr + (\mbox{\ }) x + (\mbox{\ }) y + (\mbox{\ }) P 
+ (\mbox{\ }) dP/dr, \\  
0 &= d^2P/dr^2 + (\mbox{\ }) dP/dr + (\mbox{\ }) x + (\mbox{\ }) y 
+ (\mbox{\ }) P, 
\end{align} 
\end{subequations} 
where the various coefficients $(\mbox{\ })$ are functions of $r$
constructed from the background fluid variables. The explicit form 
of these coefficients will be displayed below for the case of the
modified $n=1$ polytrope discussed in Sec.~\ref{sec:star}. For now,
it suffices to notice that an analysis of Eqs.~(\ref{xyP_system}) 
near $r=0$ reveals that when $\ell \neq 0$, the variables possess the
asymptotic behavior  
\begin{equation} \label{reg_cond}
x \sim x_0 r^{\ell-1}, \qquad 
y \sim y_0 r^\ell, \qquad 
P \sim P_0 r^\ell, 
\end{equation} 
where $x_0$ and $P_0$ are freely specifiable constants, and 
$\ell y_0 = \omega^2 (\mu_c + p_c) x_0 + \ell (\mu_c - 3 p_c) P_0$,   
with $\mu_c = \mu(r=0)$, $p_c = p(r=0)$. When $\ell = 0$ we have 
instead 
\begin{equation} 
x \sim x_1 r, \qquad 
y \sim y_0, \qquad 
P \sim P_0 + P_2 r^2,  
\end{equation} 
where $x_1$ and $P_0$ are freely specifiable, while $y_0$ and $P_2$
can be expressed in terms of them. 

The boundary conditions at $r=R$ are determined by $\Delta p = 0$,
which identifies the true position of the surface, and the requirement
that $\delta \Phi$ be smoothly joined with the external solution. The
first equation gives $\delta p + \xi_r dp/dr = 0$, or 
$y - m(\mu-3p)x/r^2 = 0$. Assuming that the background density $\mu$
goes to zero on the boundary, we find that  
\begin{equation} 
y(r=R) = 0. 
\label{y_boundary} 
\end{equation} 
Outside the body, the differential equation satisfied by $P$ becomes 
\begin{equation} 
\frac{d^2P}{dr^2} + \frac{2}{r} \frac{dP}{dr} + \biggl[ \omega^2 
- \frac{\ell(\ell+1)}{r^2} \biggr] P = 0 .
\end{equation} 
This is the spherical Bessel equation, and the
solution describing an outgoing wave is 
\begin{equation} 
P_{\rm ext} = A h_\ell^{(1)}(\omega r), 
\label{P_ext} 
\end{equation} 
where $A$ is a constant amplitude and 
$h_\ell^{(1)} = j_\ell + i y_\ell$ is a spherical Hankel function.
Equations (\ref{y_boundary}) and (\ref{P_ext}) provide the required
boundary conditions at $r=R$. 

\subsection{Modified $n=1$ polytrope} 
\label{subsec:perturbed3}

For concreteness we now choose the unperturbed configuration to be the
modified $n=1$ polytrope described in Sec.~\ref{sec:star}. 
For simplicity we set  
\begin{equation} 
\Gamma_1 = 2, \qquad \Gamma_3-1 = 0. 
\end{equation} 
The constant value for $\Gamma_1$ is a crude approximation that is
entirely sufficient for our purposes. The exact expression, assuming
that the perturbed fluid possesses the same equation of state as the
unperturbed configuration, can be obtained from Eq.~(\ref{Gamma_n1}), 
and the approximation $\Gamma_1 \simeq 2$ is accurate when $M/R \ll 1$. 
We note that $\Gamma < \Gamma_1$ everywhere within the body, which 
suggests that it is locally stable against convection (see Sec.~9.3 of 
Ref.~\cite{friedman-stergioulas:13}). 

The explicit system of perturbation equations for this model is as
follows. We first introduce dimensionless variables $\bar{r} := r/R$, 
$\chi := M/R$, and $w := \omega \sqrt{R^3/M}$. When $\ell \neq 0$ we
introduce new dimensionless variables $e_j$ such that  
\begin{equation} 
x = R \bar{r}^{\ell-1} e_1, \qquad 
y = \chi^2 R^{-2} \bar{r}^\ell e_2, \qquad 
P = \chi \bar{r}^\ell e_3, \qquad 
\frac{dP}{dr} = \chi R^{-1} \bar{r}^{\ell-1} e_4, 
\label{ej_def} 
\end{equation} 
and express the differential equations in the form 
\begin{equation} 
0 = {\cal E}_j := \bar{r} \frac{d e_j}{d\bar{r}} + {\cal A}_j^{\ k} e_k, 
\label{deq_ej} 
\end{equation} 
where summation over the repeated index $k$ is understood. The
nonvanishing components of the matrix  ${\cal A}_j^{\ k}$ are given by  
\begin{subequations} 
\begin{align} 
{\cal A}_1^{\ 1} &= \ell + \frac{\pi \bar{r} \cos \pi \bar{r}}{\sin\pi \bar{r}}, \\ 
{\cal A}_1^{\ 2} &= -\frac{ 4\pi \bar{r}^2 \bigl\{  
\bigl[ \ell(\ell+1) - 2\chi(w \bar{r})^2 \bigr] \sin \pi \bar{r} 
- \pi w^2 \bar{r}^3 \bigr\} }{ w^2  
(\pi \bar{r} + 2\chi \sin \pi \bar{r}) \sin^2 \pi \bar{r}}, \\ 
{\cal A}_1^{\ 3} &= \frac{ \ell(\ell+1) \pi \bar{r} }{ w^2 
(\pi \bar{r} + 2\chi \sin \pi \bar{r})}, \\ 
{\cal A}_2^{\ 1} &= \frac{\chi}{2 \pi^2 \bar{r}^5}\, \sin^3\pi \bar{r} 
- \chi\biggl( \frac{w^2}{2\pi \bar{r}^2} + \frac{\cos\pi \bar{r}}{\pi
 \bar{r}^4} \biggr) \sin^2\pi \bar{r} 
- \biggl( \frac{w^2}{4\bar{r}} - \frac{\chi \cos^2\pi 
\bar{r}}{2\bar{r}^3} \biggr) \sin \pi \bar{r}, \\ 
{\cal A}_2^{\ 2} &= \ell + 1 + \chi\cos\pi \bar{r} 
- \frac{\chi \sin\pi \bar{r}}{\pi \bar{r}} 
- \frac{\pi \bar{r} \cos\pi \bar{r}}{\sin\pi \bar{r}}, \\ 
{\cal A}_2^{\ 3} &= -\frac{\chi \sin^2\pi \bar{r}}{4 \pi \bar{r}^2} 
+ \frac{\chi \sin\pi \bar{r} \cos\pi \bar{r}}{4\bar{r}}, \\ 
{\cal A}_2^{\ 4} &= -\frac{\sin\pi \bar{r}}{4 \bar{r}}, \\
{\cal A}_3^{\ 3} &= \ell, \\ 
{\cal A}_3^{\ 4} &= -1, \\ 
{\cal A}_4^{\ 1} &= \frac{2 \chi \sin^2\pi \bar{r}}{\bar{r}^2} 
- \frac{2\pi \chi \sin\pi \bar{r} \cos\pi \bar{r}}{\bar{r}}, \\ 
{\cal A}_4^{\ 2} &= \frac{4\pi^2 \bar{r}^3}{\sin\pi \bar{r}} 
- 4\pi \chi \bar{r}^2, \\ 
{\cal A}_4^{\ 3} &= -\ell(\ell+1) + \chi (w \bar{r})^2 
- \chi \pi \bar{r} \sin\pi \bar{r}, \\ 
{\cal A}_4^{\ 4} &= \ell + 1. 
\end{align} 
\end{subequations} 
The variables admit the expansions 
\begin{equation} 
e_j = u_{j0} + u_{j2} \bar{r}^2 + u_{j4} \bar{r}^4 + \cdots 
\label{ej_r0} 
\end{equation} 
near $\bar{r}=0$. Analysis of Eqs.~(\ref{deq_ej}) around the origin 
reveals that of all the expansion coefficients in Eq.~(\ref{ej_r0}), 
$e_1(\bar{r}=0) = u_{10}$ and $e_3 (\bar{r}=0) = u_{30}$ are 
freely-specifiable, and the remaining ones
are determined in terms of them by the differential equations. 
In particular, regularity considerations imply that
\begin{equation} \label{bc1}
e_2 (\bar{r}=0) = u_{20} = \frac{\pi \omega^2}{4\ell} (1 + 2\chi )
 u_{10} + \frac{\pi}{4} u_{30}, \qquad 
e_4(\bar{r}=0) = u_{40} = \ell u_{30}.
\end{equation}
Near $\bar{r}=1$ we have instead 
\begin{equation} 
e_j = v_{j0} + v_{j1}(\bar{r}-1) + v_{j2}(\bar{r}-1)^2 
+ v_{j3}(\bar{r}-1)^3 + \cdots.
\label{ej_r1} 
\end{equation}
The boundary condition (\ref{y_boundary}) and the requirement that $P$ 
be smoothly joined with the external solution (\ref{P_ext}) imply
\begin{equation} \label{bc2}
e_2(\bar{r}=1) = v_{20} = 0, \qquad
e_4(\bar{r}=1) = v_{40} =  \omega R 
\frac{h^{(1)\prime}_\ell(\omega R)}{h^{(1)}_\ell(\omega R)} v_{30}
\end{equation}
where a prime indicates differentiation with respect to the argument. 
Thus, at $\bar{r}=1$ there are two freely specifiable constants, 
$v_{10}$ and $v_{30}$, and the remaining coefficients are determined
by the differential equations and by the boundary conditions 
(\ref{bc2}).

When $\ell = 0$ the new variables $e_j$ are defined by   
\begin{equation} 
x = R \bar{r} e_1, \qquad 
y = \chi^2 R^{-2} e_2, \qquad 
P = \chi e_3, \qquad 
\frac{dP}{dr} = \chi R^{-1} \bar{r} e_4, 
\end{equation} 
and the differential equations are again put in the form of
Eq.~(\ref{deq_ej}). In this case the nonvanishing components of 
${\cal A}_j^{\ k}$ are given by
\begin{subequations} 
\begin{align} 
{\cal A}_1^{\ 1} &= 2 + \frac{\pi \bar{r} \cos \pi 
\bar{r}}{\sin\pi \bar{r}}, \\ 
{\cal A}_1^{\ 2} &= \frac{4\pi \bar{r}^2}{\sin^2 \pi \bar{r}}, \\ 
{\cal A}_2^{\ 1} &= \frac{\chi}{2 \pi^2 \bar{r}^3}\, \sin^3\pi \bar{r} 
- \chi \biggl( \frac{w^2}{2\pi} + \frac{\cos\pi \bar{r}}{\pi \bar{r}^2} 
\biggr) \sin^2\pi \bar{r} 
- \biggl( \frac{w^2 \bar{r}}{4} - \frac{\chi \cos^2\pi 
\bar{r}}{2\bar{r}} \biggr) \sin \pi \bar{r}, \\ 
{\cal A}_2^{\ 2} &= 1 + \chi \cos\pi \bar{r} 
- \frac{\chi \sin\pi \bar{r}}{\pi \bar{r}} 
- \frac{\pi \bar{r} \cos\pi \bar{r}}{\sin\pi \bar{r}}, \\ 
{\cal A}_2^{\ 3} &= -\frac{\chi \sin^2\pi \bar{r}}{4 \pi \bar{r}^2} 
+ \frac{\chi \sin\pi \bar{r} \cos\pi \bar{r}}{4\bar{r}}, \\ 
{\cal A}_2^{\ 4} &= -\frac{1}{4} \bar{r} \sin\pi \bar{r}, \\
{\cal A}_3^{\ 4} &= -\bar{r}^2, \\ 
{\cal A}_4^{\ 1} &= \frac{2\chi\sin^2\pi \bar{r}}{\bar{r}^2} 
- \frac{2\pi \chi \sin\pi \bar{r} \cos\pi \bar{r}}{\bar{r}}, \\ 
{\cal A}_4^{\ 2} &= \frac{4\pi^2 \bar{r}}{\sin\pi \bar{r}} 
- 4\pi \chi, \\ 
{\cal A}_4^{\ 3} &= \chi w^2 - \frac{\pi M}{\bar{r}} \sin\pi\bar{r}, \\ 
{\cal A}_4^{\ 4} &= 3. 
\end{align} 
\end{subequations} 
In this case also the variables admit the expansions of
Eqs.~(\ref{ej_r0}) and (\ref{ej_r1}), with $u_{10}$, $u_{30}$, 
$v_{10}$, and $v_{30}$ freely specifiable and the remaining coefficients
determined by the differential equations and by the boundary conditions
(\ref{bc2}) with $\ell = 0$.

\subsection{Eigenfrequencies and quasinormal modes} 
\label{subsec:eigen} 

Equations (\ref{deq_ej}) form a fourth-order system of differential
equations. In principle, the set of regularity and boundary conditions 
given in Eqs.~(\ref{bc1}) and (\ref{bc2}) provides the required
data to close the system. However, since Eqs.~(\ref{deq_ej}) are
linear and homogeneous, the functions $e_j$ are determined only up to a
multiplicative constant; this implies that the system will have no 
solution unless $\omega$ is chosen appropriately. The system is
therefore an eigenvalue problem for $\omega$. Because $P_{\rm ext}$
is necessarily a complex function, the solution to the eigenvalue
problem is a set of complex frequencies $\omega := \sigma -i \kappa$,where $\sigma$ and $\kappa$ are both real. 

The eigenfrequencies are determined as follows. The differential 
equations (\ref{deq_ej}) for $e_j(r)$ are first integrated 
outward from $\bar{r}=0$, together with (\ref{bc1}), and with $u_{10}$ 
chosen arbitrarily to provide a  normalization for the solution. Next 
the differential equations are integrated inward from $\bar{r}=1$, 
together with Eq.~(\ref{bc2}). There are four freely specifiable 
complex constants, $\{ \omega, u_{30}, v_{10}, v_{30} \}$ or
$\{ \omega, u_{30}, v_{10}, A \}$, 
and these are adjusted so that at a point $\bar{r} = r_1$ such that 
$0 < r_1 < 1$, the functions $e^{\rm out}_j(r)$, integrated outward
from $\bar{r}=0$, match the functions $e^{\rm in}_j(r)$,
integrated inward from $\bar{r}=1$: this provides a set of four
(complex) conditions to be solved for the four unknown constants. The
search for the solution is performed with a globally convergent
version of the Newton-Raphson method, as described in Sec.~9.7 of  
Ref.~\cite{numerical-recipes:02}. 
The numerical computation of high overtones (i.e., with large 
values of $|n|$; see below) is particularly challenging, since 
$|\Im(\omega)|$ decays exponentially with increasing $|n|$, as does 
the constant $A$. An increasingly high resolution is required in order
to compute them, and we find that floating-point operations with 
double precision are not always enough to achieve convergence. We
overcome this technical issue by computing these high overtones in
both \textit{Mathematica} and \textit{Maple}, in which a higher 
numerical precision can be easily stipulated.

Once a solution is found for $\sigma$ (the mode frequency) and
$\kappa$ (the inverse of the damping time), the perturbation 
equations can be integrated once more to obtain 
the perturbation variables $\xi_r = x(r) Y_{\ell m}(\theta,\phi) 
e^{-\kappa t} e^{i\sigma t}$, $\delta p = y(r) Y_{\ell m}(\theta,\phi) 
e^{-\kappa t} e^{i\sigma t}$, and $\delta \Phi 
= P(r) Y_{\ell m}(\theta,\phi) e^{-\kappa t} e^{i\sigma t}$ throughout
the fluid; these describe the quasinormal modes of the body. For each
$\ell$ the modes are characterized by the number $n$ of nodes in
$x(r)$ or $y(r)$. The mode with $n=0$ is the {\it fundamental mode}
($f$-mode), and modes with $n > 0$ divide themselves into 
{\it pressure modes} ($p$-modes) and {\it gravity modes}
($g$-modes). The frequencies of the $p$-modes are higher than the
$f$-mode frequency, and they increase with the mode number $n$. On the
other hand, the frequencies of the $g$-modes are lower than the
$f$-mode frequency, and they decrease with increasing $|n|$. It is
conventional to assign a negative value of $n$ to the $g$-modes, and
we adopt this practice here. 

\begin{figure} 
\includegraphics[width=1.0\linewidth]{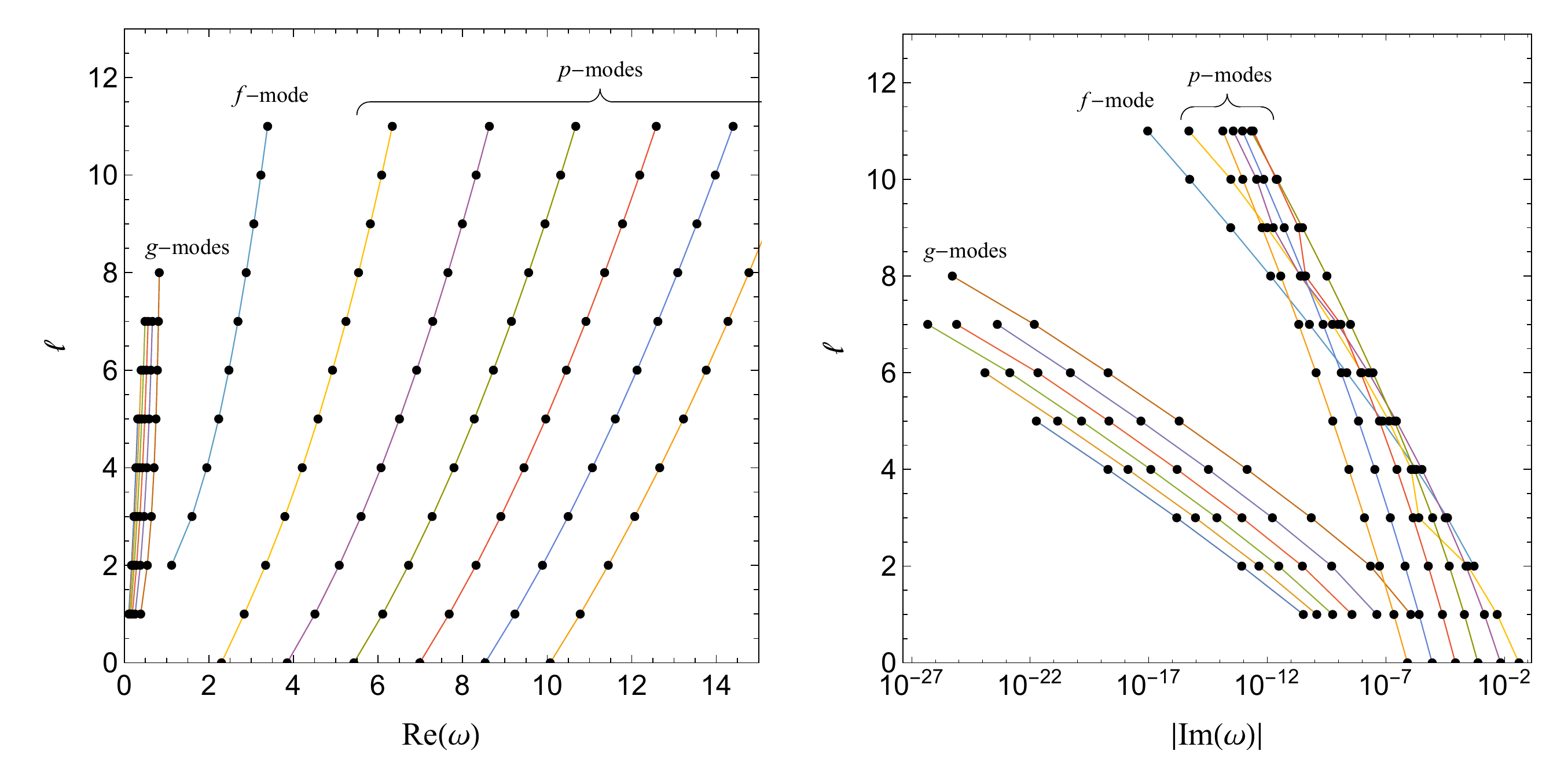}
\caption{Real and (modulus of the) imaginary parts of the 
  quasinormal-mode frequencies $\omega$ of a fluid body with compactness
  $M/R = 0.3$ in the scalar theory of gravity, for the lower multipole
  orders. Frequencies are presented in units of $\sqrt{M/R^3}$. The mode
  frequencies are grouped in families of definite mode number $n$, and
  classified as $g$-, $f$- or $p$-modes as described in the main text.}  
\label{fig:fig7} 
\end{figure} 

The result of our computations for a fluid body with $\chi = 0.3$
were already presented in Fig.~\ref{fig:fig1}, and another perspective
is offered in Fig.~\ref{fig:fig7}. It was pointed out that many of the
quasinormal modes, in particular the $g$-modes, have extremely small
values of $\kappa$, which makes them extremely long-lived. 
It is also noteworthy that a number of $p$-modes are in fact unstable,
with $\kappa < 0$. These unstable modes, however, have virtually no
impact on the self-force results presented in Sec.~\ref{sec:particle}. 

\subsection{Static perturbations} 
\label{subsec:static} 

The case $\omega = 0$ requires a separate treatment. In this
case the manipulations leading to the system of equations
(\ref{xyP_system}) break down, because as Eq.~(\ref{y_vs_z})
indicates, $y$ is no longer independent from $P$. A further
study of the perturbation equations also reveals that $x = (r^2/m) P$,  
and that $z$ can be expressed in terms of $P$ and its first derivative.
The entire content of the perturbation equations is therefore 
summarized in the equation for $P$, which becomes 
\begin{equation} 
r^2 \frac{d^2P}{dr^2} + 2r \frac{dP}{dr} - \biggl[ 12\pi r^2 (\mu-3p) 
+ \frac{4\pi r^4}{m} \frac{d\mu}{dr} + \ell(\ell+1) \biggr] P = 0. 
\end{equation} 
When the unperturbed configuration is that of a modified $n=1$
polytrope, the differential equation becomes 
\begin{equation} 
r^2 \frac{d^2P}{dr^2} + 2r \frac{dP}{dr} + \bigl[ (\pi r/R)^2 - \ell(\ell+1) \bigr] P = 0.  
\end{equation} 
This is the spherical Bessel equation, with solution 
\begin{equation} 
P = B j_\ell(\pi r/R),  
\label{P_static} 
\end{equation} 
where $B$ is an arbitrary amplitude. 

\section{Scalar gravity with fluid body and point particle}  
\label{sec:particle} 

In this section we generalize the foregoing discussion by allowing the
system to also include a point particle, which is assumed to move
outside the fluid body on a world line described by the parametric
relations $z^\alpha(s)$, where $s$ is proper time on the world
line. We continue to set $\alpha(\Phi) = 1$ and to take the spacetime
metric to be flat.   

\subsection{Defining equations} 

The insertion of the point particle is accomplished by generalizing
the field equations to 
\begin{equation} 
\Box \Phi = 4\pi \bigl( T^\mu_{\ \mu} 
+ \mathfrak{t}^\mu_{\ \mu} \bigr), \qquad 
\nabla_\beta \bigl( T^{\alpha\beta} + \mathfrak{t}^{\alpha\beta} \bigr) 
= -\bigl( T^\mu_{\ \mu} 
+ \mathfrak{t}^\mu_{\ \mu} \bigr) \nabla^\alpha \Phi, 
\end{equation} 
where $T^{\alpha\beta}$ continues to denote the fluid's
energy-momentum tensor, while
\begin{equation} 
\mathfrak{t^{\alpha\beta}} = \int \m\, v^\alpha v^\beta 
\delta_4(x,z)\, ds 
\end{equation} 
is the energy-momentum tensor of the particle, with $\m(s)$ denoting
its (time-dependent) mass, $v^\alpha = dz^\alpha/ds$ its velocity vector, 
and $\delta_4(x,z)$ a scalarized Dirac delta-function supported on the
particle's world line --- $x$ stands for an arbitrary spacetime point,
while $z(s)$ is a point on the world line. By virtue of the assumption
that the particle moves outside the fluid distribution, the supports
of $T^{\alpha\beta}$ and $\mathfrak{t}^{\alpha\beta}$ are disjoint,
and the equations of motion decompose into a conservation equation
for the fluid alone, 
\begin{equation} 
\nabla_\beta T^{\alpha\beta} = -T^\mu_{\ \mu}\, \nabla^\alpha \Phi,  
\end{equation} 
and a conservation equation for the particle alone, 
\begin{equation} 
\nabla_\beta \mathfrak{t}^{\alpha\beta}  
= -\mathfrak{t}^\mu_{\ \mu}\, \nabla^\alpha \Phi. 
\end{equation} 
The first equation produces the same fluid equations that were studied
in the preceding sections. The second equation implies 
\begin{equation} 
\frac{D}{ds} \bigl( \m v^\alpha \bigr) = \m \nabla^\alpha \Phi(z), 
\end{equation} 
which can be decomposed into 
\begin{equation} 
\frac{d\m}{ds} = - \m v^\alpha \nabla_\alpha \Phi(z), \qquad 
\frac{Dv^\alpha}{ds} = \bigl( \eta^{\alpha\beta} 
+ v^\alpha v^\beta \bigr) \nabla_\beta \Phi(z). 
\label{eom} 
\end{equation} 
The particle and the fluid are coupled through the potential $\Phi$,
which now satisfies the wave equation  
\begin{equation} 
\Box \Phi = -4\pi (\mu - 3p) 
- 4\pi \int \m\, \delta_4(x,z)\, ds. 
\label{BoxPhi} 
\end{equation} 
In the body's interior the second term vanishes, and the wave equation
reduces to the one studied in the preceding sections. In the exterior
the first term vanishes, and $\Phi$ is sourced by the particle only.     

We observe that the insertion of the particle has no impact on the
fluid equations, which are identical to those considered
previously. The particle nevertheless influences the fluid's  
dynamics through the boundary values adopted by $\Phi$, which
are now different from those considered previously. The particle
is also affected by the fluid, through the same boundary conditions. The
motion of the fluid and the motion of the particle are inherently
coupled.  

\subsection{Unperturbed and perturbed configurations} 

As in Sec.~\ref{sec:perturbed} we take the unperturbed configuration
of the system to be a spherically-symmetric distribution of fluid in
hydrostatic equilibrium, in the absence of a point particle. This
configuration is described by the unperturbed variables $\mu$, $p$,
$u^\alpha$, and $\Phi$; outside the body $\Phi = M/r$. The insertion
of the particle creates a perturbed configuration, and the
perturbation is described by the variables $\Delta \mu$, $\Delta p$,
$\Delta u^\alpha$, $\delta \Phi$, $\m$, and $v^\alpha$.  

As we observed at the end of the preceding subsection, the equations
for the fluid perturbations take the same form as those studied
previously. On the other hand, the equation for $\delta \Phi$ becomes  
\begin{equation} 
\Box \delta \Phi = -4\pi (\delta \mu - 3\delta p) 
- 4\pi \int \m\, \delta_4(x,z)\, ds,  
\end{equation} 
which reduces to Eq.~(\ref{delta_wave}) inside the body, and to 
\begin{equation} 
\Box \delta \Phi_{\rm out} = -4\pi \int \m\, \delta_4(x,z)\, ds  
\label{BoxPhi_out} 
\end{equation} 
outside the body. The particle's variables are determined by 
\begin{equation} 
\frac{d\m}{ds} = -\m v^\alpha \nabla_\alpha \Phi 
- \m v^\alpha \nabla_\alpha \delta \Phi_{\rm out} 
\label{dmds_full} 
\end{equation} 
and 
\begin{equation} 
\frac{Dv^\alpha}{ds} = \bigl( \eta^{\alpha\beta} + v^\alpha v^\beta
\bigr) \nabla_\beta \Phi
+ \bigl( \eta^{\alpha\beta} + v^\alpha v^\beta
\bigr) \nabla_\beta \delta \Phi_{\rm out},  
\label{Dvds_full} 
\end{equation} 
with the first set of terms representing the particle's motion in the 
background potential $\Phi$, and the second set representing the
self-force. 

We take the motion of the particle in the background potential 
$\Phi = M/r$ to be a circular orbit of radius $r = r_0$ and angular
velocity $\Omega$ in the equatorial plane ($\theta = \pi/2$) of the
fluid body. A simple calculation reveals that in the coordinates $(t,
r, \theta, \phi)$, the components of the velocity vector are given by   
\begin{equation} 
v^\alpha = \gamma (1, 0, 0, \Omega ), \qquad 
\gamma = \sqrt{1+M/r_0}, \qquad 
\Omega^2 = \frac{M/r_0^3}{1 + M/r_0}.
\label{v_def}  
\end{equation} 
Because $v^\alpha \nabla_\alpha \Phi = 0$, the particle's mass $\m$ is
conserved in the background motion. 

The particle creates the perturbation $\delta \Phi_{\rm out}$, and for 
the circular orbit the source term of Eq.~(\ref{BoxPhi_out}) can be
written as 
\begin{align} 
4\pi \int \m\, \delta_4(x,z)\, ds &= 
\frac{4\pi\m}{\gamma r_0^2} \delta(r-r_0) \delta(\theta-\pi/2)
\delta(\phi - \Omega t) 
\nonumber \\ &=
\frac{4\pi\m}{\gamma r_0^2} \delta(r-r_0) \sum_{\ell m}
Y^*_{\ell m}(\pi/2,\Omega t) Y_{\ell m}(\theta, \phi) 
\nonumber \\ &=
\frac{4\pi\m}{\gamma r_0^2} \delta(r-r_0) \sum_{\ell m}
e^{-i m \Omega t} Y_{\ell m}(\pi/2,0) Y_{\ell m}(\theta, \phi),  
\end{align} 
where an asterisk indicates complex conjugation, and the sums over
$\ell$ and $m$ run from $0$ to $\infty$ and from $-\ell$ to $\ell$,
respectively. 

The preceding expression indicates that the mass density has a
frequency spectrum limited to harmonics of the orbital frequency
$\Omega$. It implies that $\delta \Phi_{\rm out}$ can be expanded as 
\begin{equation} 
\delta \Phi_{\rm out} = \sum_{\ell m} P_{\ell m}(r) 
Y_{\ell m}(\theta,\phi) e^{-i m \Omega t}, 
\label{deltaPhi_out} 
\end{equation} 
with the radial functions $P_{\ell m}(r)$ determined by
Eq.~(\ref{BoxPhi_out}). The radial equation takes the form of  
\begin{equation} 
\frac{d^2 P_{\ell m}}{dr^2} + \frac{2}{r} \frac{d P_{\ell m}}{dr} 
+ \biggl[ \omega^2 - \frac{\ell(\ell+1)}{r^2} \biggr] P_{\ell m} 
= -\frac{G_{\ell m}}{r_0^2} \delta(r-r_0), 
\label{Pdeq}
\end{equation} 
where $\omega = m\Omega$ and
\begin{equation}
G_{\ell m} := \frac{4\pi \m}{\gamma}Y_{\ell m}(\pi/2,0).      
\label{Glm} 
\end{equation} 
It should be noted that the sum over $m$ in Eq.~(\ref{deltaPhi_out})
is restricted to values such that $\ell + m$ is even, because 
$Y_{\ell m}(\pi/2,0)$ vanishes when $\ell + m$ is odd. For $\ell + m$
even we have instead 
\begin{equation} 
Y_{\ell m}(\pi/2,0)  = (-1)^{\frac{1}{2}(\ell + m)} 
\sqrt{\frac{2\ell+1}{4\pi}} 
\frac{\sqrt{(\ell-m)!(\ell+m)!}}{(\ell-m)!! (\ell+m)!!}. 
\label{Ylm} 
\end{equation} 
It should also be noted that $\delta \Phi_{\rm out}$ is real when  
$P_{\ell m}$ satisfies the conditions 
\begin{equation} 
P_{\ell,-m} = (-1)^m P_{\ell m}^*. 
\label{P_reality} 
\end{equation} 
These conditions have to be respected when boundary conditions are
adopted for the radial functions. 

\subsection{Solution to the perturbation equations ($m \neq 0$)} 
\label{subsec:sol_general} 

We now construct the appropriate solutions to Eq.~(\ref{Pdeq}) and the
remaining perturbation equations, assuming first that $m \neq 0$, so
that $\omega = m \Omega$ is nonzero.  When $r \neq r_0$ the
differential equation reduces to the spherical Bessel equation, with
$j_\ell(\omega r)$ and $y_\ell(\omega r)$ as linearly independent
solutions. When $r > r_0$ the solution must represent an outgoing
wave, and to reflect this we write  
\begin{equation} 
P^>_{\ell m} = \omega G_{\ell m} A_\ell(\omega) 
\bigl[ j_\ell(\omega r) + i y_\ell(\omega r) \bigr], 
\end{equation} 
where $A_\ell(\omega)$ is a complex amplitude and a factor of 
$\omega G_{\ell m}$ was inserted for convenience. For $r < r_0$ we
have instead 
\begin{equation} 
P^<_{\ell m} = \omega G_{\ell m} \bigl[ B_\ell(\omega) 
j_\ell(\omega r) + C_\ell(\omega) y_\ell(\omega r) \bigr],  
\end{equation} 
where $B_\ell(\omega)$ and $C_\ell(\omega)$ can be related to
$A_\ell(\omega)$ by the junction conditions implied by the presence of
a delta-function on the right-hand side of Eq.~(\ref{Pdeq}). A simple
calculation returns 
\begin{equation} 
B_\ell(\omega) = A_\ell(\omega) - y_\ell(\omega r_0), \qquad 
C_\ell(\omega) = i A_\ell(\omega) + j_\ell(\omega r_0); 
\end{equation} 
the end result was simplified by exploiting the Wronskian relation 
$j_\ell y'_\ell - j'_\ell y_\ell = (\omega r)^{-2}$ satisfied by the
spherical Bessel functions, with a prime indicating differentiation
with respect to the argument $\omega r$. 

The function $P^<_{\ell m}(r)$ must be matched to a solution
of the fluid perturbation equations (\ref{deq_ej}) at the body's 
boundary, situated at $r = R$. With $\omega$ now real, these fluid
perturbation equations admit real solutions. We take advantage of this
property through the following procedure. Ignoring for the moment the 
boundary condition coming from the matching to the external potential,
we have five freely specifiable constants: 
$\{u_{10},u_{30},v_{10},v_{30},v_{40}\}$, corresponding to 
$\{e_1(r=0),e_3(r=0),e_1(r=R),e_3(r=R),e_4(r=R)\}$. 
If we fix the normalization 
by setting, e.g., $v_{10}=1$, then the four remaining constants can
be determining by adapting the method described in 
Sec.~\ref{subsec:eigen} to find the quasinormal modes of the fluid
distribution. The system of equations (\ref{deq_ej}) is integrated
outward from $r=0$ with unknown boundary values $u_{10}$ and $u_{30}$,
and it is integrated inward from $r=R$ with unknown boundary values
$v_{30}$ and $v_{40}$. We then determine the four unknown constants
by imposing continuity of $e_j(r)$ at a middle point
$r=r_1$. Let us call the solution thus obtained $\hat{e}_j(r)$.
Then, the actual complex solution, that matches $P^<_{\ell m}(r)$ at
$r=R$, can be obtained from $\hat{e}_j(r)$ by a simple (complex)
rescaling: $e_j(r) = N_\ell(\omega) \hat{e}_j(r)$.
Recalling the definitions of Eq.~(\ref{ej_def}), we
have that the matching conditions at $r=R$ are    
\begin{subequations} 
\begin{align} 
\chi N_\ell(\omega) \hat{e}_3(R) &= P^<_{\ell m}(r=R)
= \omega G_{\ell m} \Bigl\{ 
\bigl[ A_\ell(\omega) - y_\ell(\omega r_0) \bigr] j_\ell(\omega R) 
+ \bigl[ iA_\ell(\omega) +  j_\ell(\omega r_0) \bigr] y_\ell(\omega R) 
\Bigr\}, \\ 
R^{-1} \chi N_\ell(\omega) \hat{e}_4(R) &= \frac{dP^<_{\ell m}}{dr}(r=R)
= \omega^2 G_{\ell m} \Bigl\{ 
\bigl[ A_\ell(\omega) - y_\ell(\omega r_0) \bigr] j'_\ell(\omega R) 
+ \bigl[ iA_\ell(\omega) +  j_\ell(\omega r_0) \bigr] y'_\ell(\omega R) 
\Bigr\}.
\end{align} 
\end{subequations} 
These equations determine $A_\ell(\omega)$ and
$N_\ell(\omega)$ in terms of the known amplitudes $\hat{e}_3(R)$ and 
$\hat{e}_4(R)$. Writing  
\begin{equation} 
A_\ell(\omega) = \alpha_\ell(\omega) + i \beta_\ell(\omega),  
\end{equation} 
we find that 
\begin{subequations} 
\label{alpha_beta} 
\begin{align} 
\alpha_\ell(\omega) &= 
\frac{ \bigl[ R \omega (y_0 j' - j_0 y')\hat{e}_3 + (j_0 y - y_0 j)
  \hat{e}_4 \bigl] (R \omega j' \hat{e}_3 - j \hat{e}_4 )} 
{R^2\omega^2 (j'^2 + y'^2) \hat{e}_3^2 - 2R\omega (y' y + j' j) \hat{e}_3
  \hat{e}_4 + (j^2 + y^2) \hat{e}_4^2 }, \\ 
\beta_\ell(\omega) &= 
-\frac{ \bigl[ R\omega (y_0 j' - j_0 y')\hat{e}_3 + (j_0 y - y_0 j)
  \hat{e}_4 \bigl] ( R\omega y' \hat{e}_3 - y \hat{e}_4 )} 
{R^2 \omega^2 (j'^2 + y'^2) \hat{e}_3^2 - 2R\omega (y' y + j' j) \hat{e}_3
  \hat{e}_4 + (j^2 + y^2) \hat{e}_4^2 },
\end{align} 
\end{subequations} 
where $\hat{e}_3 := \hat{e}_3(R)$, $\hat{e}_4 := \hat{e}_4(R)$, 
$j_0 := j_\ell(\omega r_0)$, $y_0 := y_\ell(\omega r_0)$,
$j := j_\ell(\omega R)$, and $y := y_\ell(\omega R)$. 

\subsection{Solution to the perturbation equations ($m = 0$)} 
\label{subsec:sol_m0} 

We next  turn to the special case $m=0$, which implies 
$\omega = 0$. In this case the linearly independent solutions to
Eq.~(\ref{Pdeq}) are $r^\ell$ and $r^{-(\ell+1)}$. The appropriate
solution for $r > r_0$ is 
\begin{equation} 
P^>_{\ell 0} =\frac{Z_\ell}{2\ell+1} \frac{G_{\ell 0}}{r_0} 
 (r_0/r)^{\ell + 1}, 
\end{equation} 
where $Z_\ell$ is a real amplitude. The junction conditions at $r=r_0$
imply that  
\begin{equation} 
P^<_{\ell 0} = \frac{1}{2\ell+1} \frac{G_{\ell 0}}{r_0} \Bigl[ 
(r/r_0)^\ell + (Z_\ell - 1) (r_0/r)^{\ell + 1} \Bigr] 
\end{equation}  
is the correct solution for $r < r_0$. 

The function $P^<_{\ell 0}$ must be matched to the internal solution
of Eq.~(\ref{P_static}) at the boundary $r=R$, and the procedure
returns 
\begin{equation} 
Z_\ell-1 = \frac{1}{\bar{r}_0^{2\ell+1}} 
\frac{ \ell j_\ell(\pi) - \pi j_\ell'(\pi) }
{ (\ell+1) j_\ell(\pi) + \pi j_\ell'(\pi) }, 
\label{Z_expression} 
\end{equation} 
where $\bar{r}_0:= r_0/R$. 

When $\ell = 0$ the function $P^<_{00}$ must be constant, because a
term proportional to $r^{-1}$ would represent an unphysical shift
in the body's mass. Removing this term amounts to setting $Z_0 = 1$,
and we obtain   
\begin{equation} 
P^>_{00} = \frac{G_{00}}{r}, \qquad 
P^<_{00} = \frac{G_{00}}{r_0}. 
\end{equation} 
With $G_{00} = \sqrt{4\pi} \m/\gamma$, we find that the $\ell = m = 0$
contribution to $\delta \Phi_{\rm out}$ is given by $\m/(\gamma r)$
when $r > r_0$, which indicates that $\m/\gamma$ is the particle's
gravitational mass. The fact that $P_{00}$ is constant when $r < r_0$
implies that the $\ell = m = 0$ mode of the perturbation vanishes
inside the body.   

\subsection{Solution to the perturbation equations in the absence of a
star}  
\label{subsec:sol_nostar} 

For the eventual purpose of regularizing the mode sum for the
self-force, we examine once more a particle of mass $\m$ 
on a circular orbit of radius $r_0$ and angular velocity
$\Omega$, but this time in the absence of a fluid body.  (In this case
the particle is maintained on its orbit by an external agent instead
of the body's gravitational field.) The solutions to Eq.~(\ref{Pdeq})
are unchanged for $r > r_0$, but the solutions for $r < r_0$ must now
be nonsingular at $r=0$.  

When $m \neq 0$ the regularity condition implies that $C_\ell(\omega)
= 0$, which produces 
\begin{equation} 
P^>_{\ell m} = i \omega G_{\ell m} 
j_\ell(\omega r_0) h_\ell^{(1)}(\omega r), \qquad 
P^<_{\ell m} = i \omega G_{\ell m} 
h_\ell^{(1)}(\omega r_0) j_\ell(\omega r),  
\end{equation} 
where $\omega = m\Omega$. When $m = 0$ the regularity condition
implies that $Z_\ell = 1$, so that 
\begin{equation} 
P^>_{\ell 0} = \frac{1}{2\ell+1} \frac{G_{\ell 0}}{r_0} 
(r_0/r)^{\ell + 1},  \qquad  
P^<_{\ell 0} = \frac{1}{2\ell+1} \frac{G_{\ell 0}}{r_0} 
(r/r_0)^\ell. 
\end{equation} 
For $\ell = m = 0$ the solution is 
\begin{equation} 
P^>_{00} = \frac{G_{00}}{r}, \qquad 
P^<_{00} = \frac{G_{00}}{r_0}, 
\end{equation} 
unchanged relative to the previous expressions. 

\subsection{Self-force} 

\begin{figure} 
\includegraphics[width=1.0\linewidth]{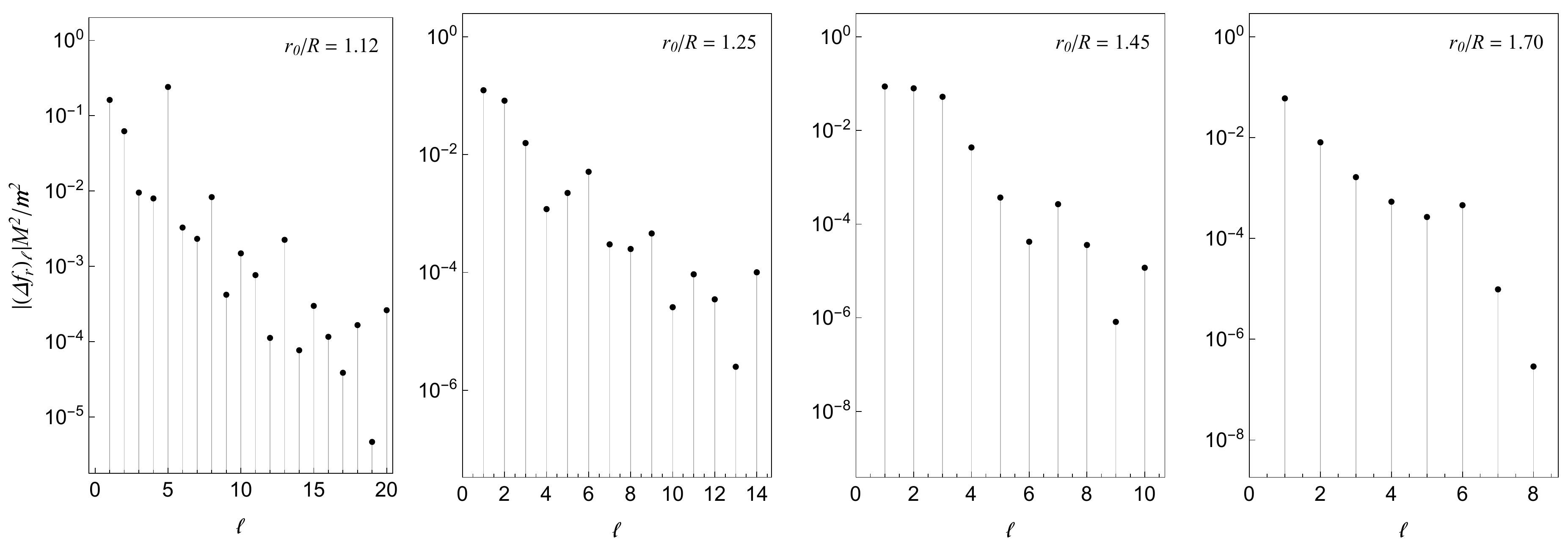}
\includegraphics[width=1.0\linewidth]{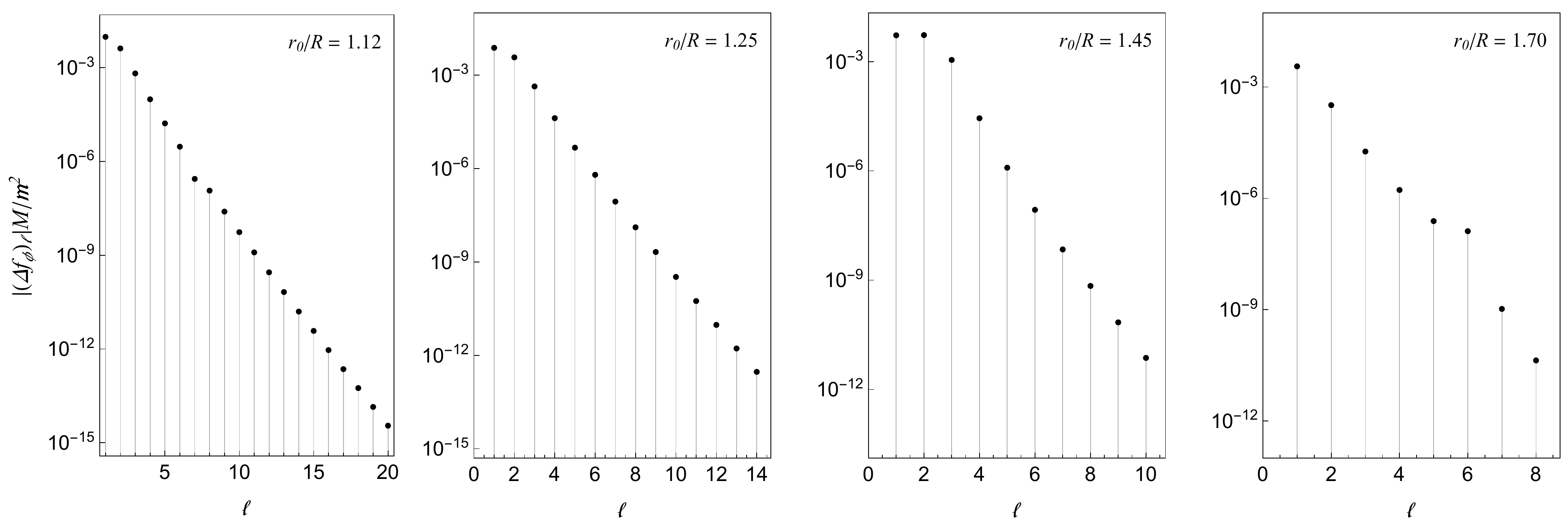}
\caption{In the upper panel the contributions of 
  $(M^2/\m^2)(\Delta\f_r)_\ell$ from low multipole orders $\ell$ are
  shown, for selected values of $r_0/R$; for each one, the sum
  from $m = -\ell$ to $\ell$ was evaluated. As a general rule, to
  achieve a specified degree of accuracy (say $10^{-5}$), the sum over 
  $\ell$ must be truncated at a larger value when $r_0/R$ is smaller.
  The rule is complicated by the resonant features, which can make a
  given contribution abnormally large. The lower panel shows the
  contributions of $(M/\m^2) (\Delta\f_\phi)_\ell$ 
  from low multipole orders $\ell$, for the same values of $r_0/R$. Here 
  the convergence of the mode sum is much faster, and is not strongly 
  affected by the resonant features.}  
\label{fig:fig8} 
\end{figure}

\begin{figure} 
\includegraphics[width=0.95\linewidth]{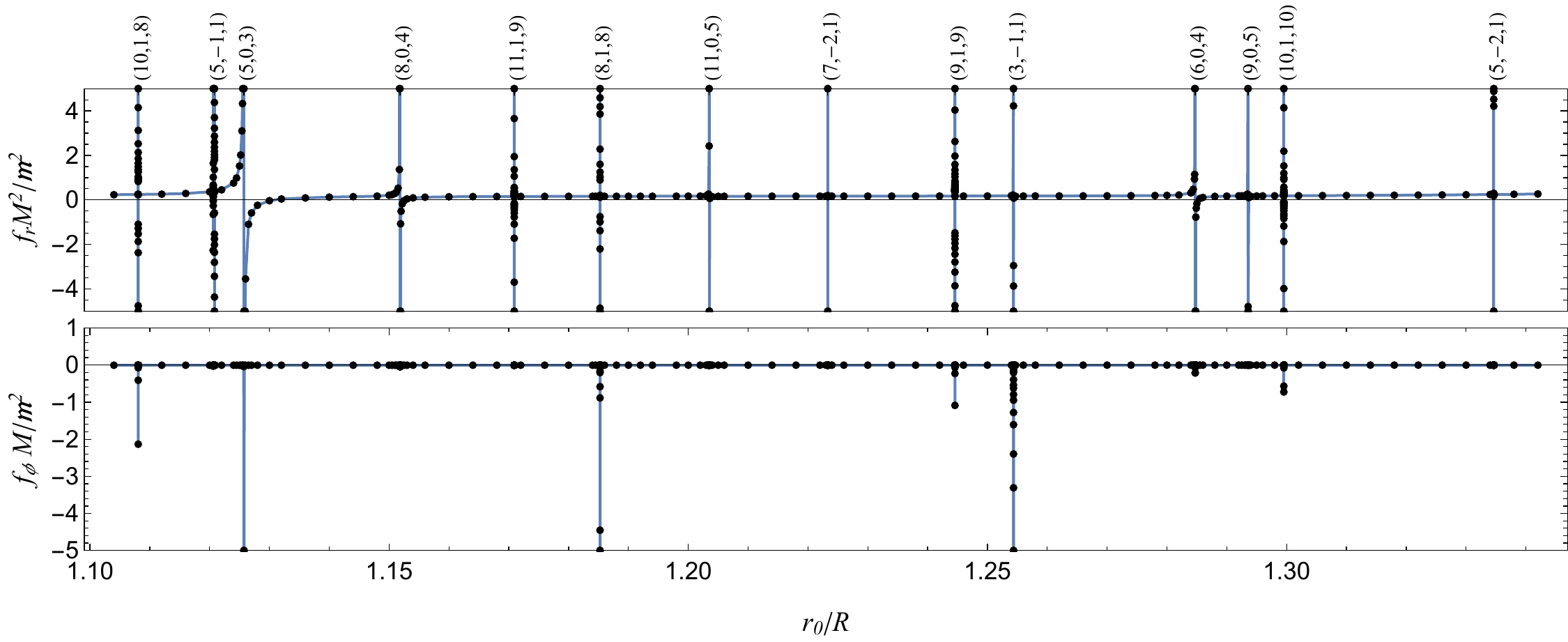}
\includegraphics[width=0.95\linewidth]{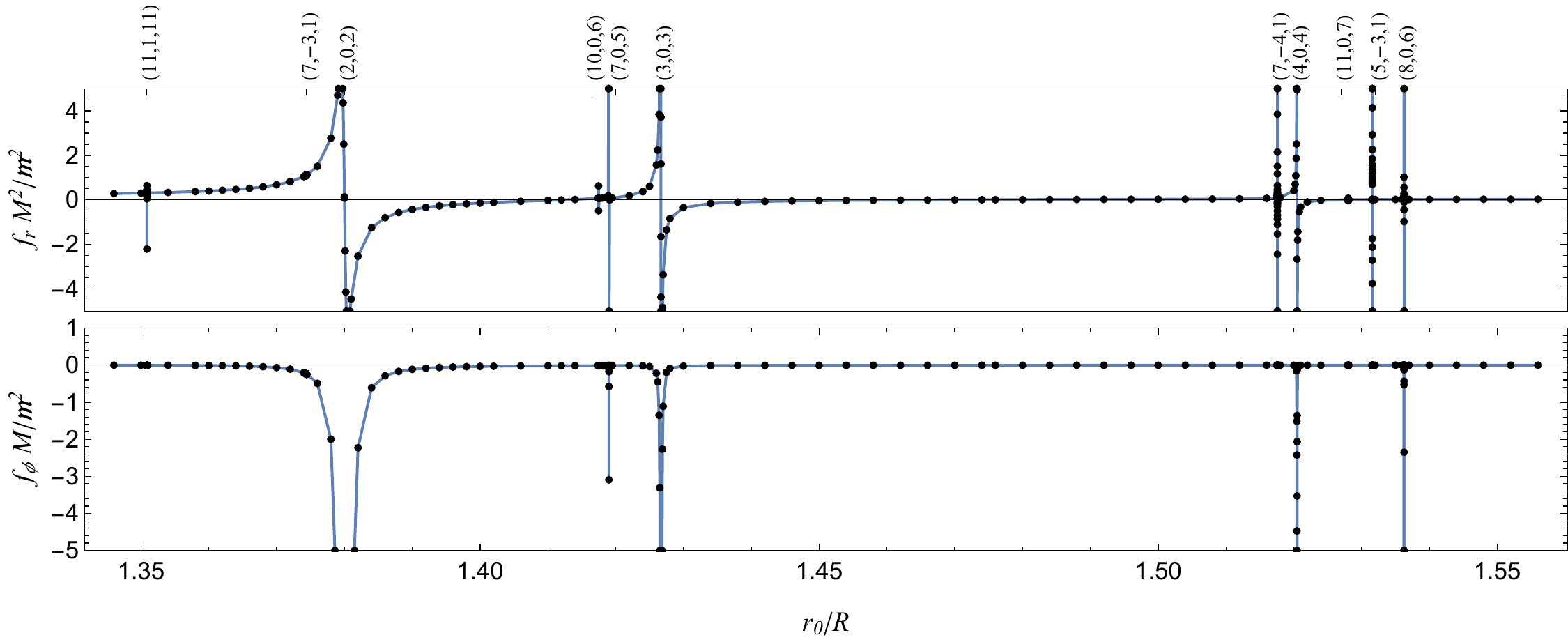}
\includegraphics[width=0.95\linewidth]{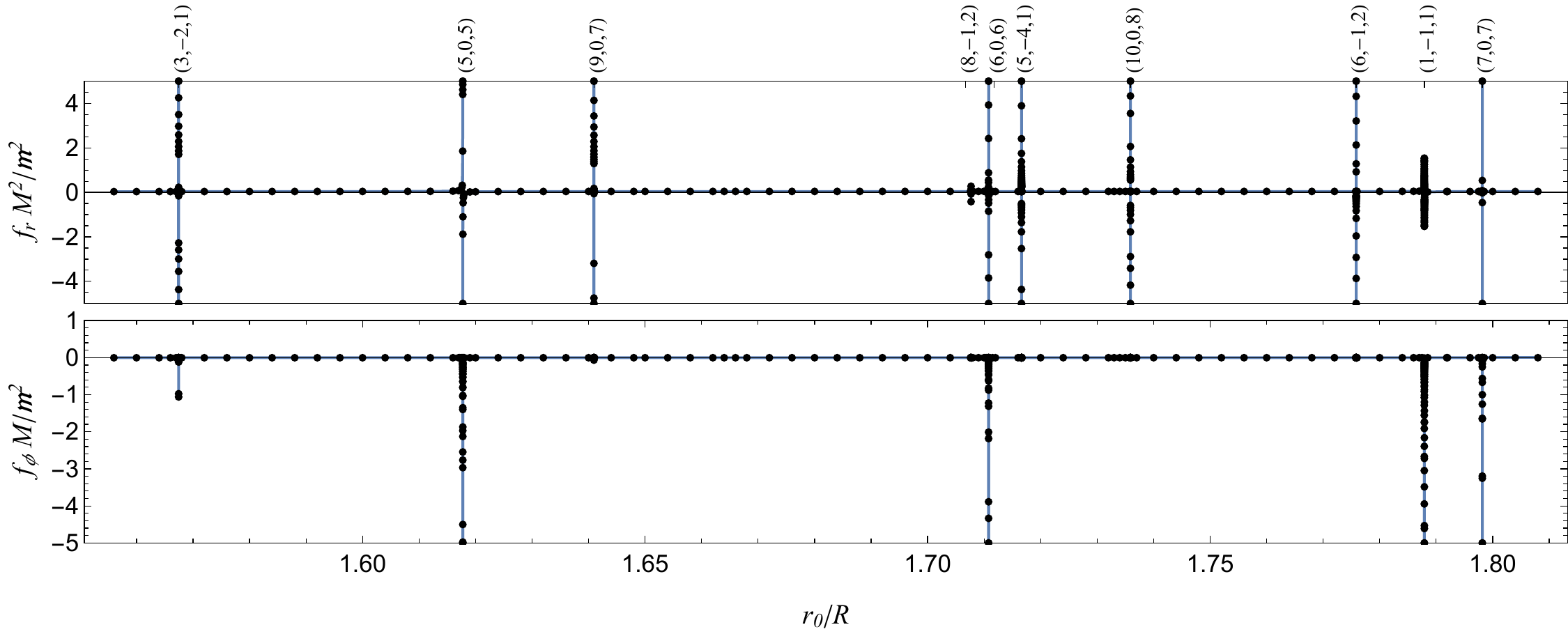}
\caption{Plots of $(M^2/\m^2) \f_r$ (upper panels) and $(M/\m^2)
  \f_\phi$ (lower panels) as functions of $r_0/R$ in three successive
  subintervals of $1.10 < r_0/R < 1.81$. Each resonant feature in the
  self-force is associated with the relevant quasinormal mode, labeled
  by $(\ell,n,m)$. (The resonance identified by $(5,-1,1)$ actually
  consists of two close resonant peaks, corresponding to $(5,-1,1)$ and
  $(7,1,7)$.) Note that while quasinormal modes are independent of
  $m$, so that only $\ell$ and $n$ are true mode labels, a resonance
  is achieved when $m \Omega = \Re(\omega)$, so that $m$ is a required  
  label for each resonance. Notice, in particular, the broad resonance
  near $r_0/M \simeq 1.38$, which is associated with the $\ell = 2$ 
  $f$-mode ($n=0$), and requires $m=2$ for a frequency match.}   
\label{fig:fig9} 
\end{figure} 

The self-force acting on the particle is described by the second set
of terms in Eqs.~(\ref{dmds_full}) and (\ref{Dvds_full}), 
\begin{equation} 
\dot{\m} := \frac{d\m}{ds} = -\m v^\alpha \nabla_\alpha 
\delta \Phi  
\label{mdot} 
\end{equation} 
and 
\begin{equation} 
\f^\alpha := \m \frac{Dv^\alpha}{ds} = \m \bigl( \eta^{\alpha\beta}  
+ v^\alpha v^\beta \bigr) \nabla_\beta \delta \Phi, 
\label{sf_def} 
\end{equation} 
where $v^\alpha$ is the velocity vector of Eq.~(\ref{v_def}), and
$\delta \Phi \equiv \delta \Phi_{\rm out}$ is the potential of
Eq.~(\ref{deltaPhi_out}), which in unbounded in the limit 
$x \to z(s)$, that is, when evaluated at the position of the particle,
$r = r_0$, $\theta = \pi/2$, and $\phi = \Omega t$.  

To regularize the expressions of Eqs.~(\ref{mdot}) and (\ref{sf_def})
we examine the {\it difference} between two self-forces 
\cite{drivas-gralla:11}, the first resulting in the presence of the
fluid body, the other resulting in its absence; in both cases the
particle moves on the same circular orbit of radius $r_0$ and angular 
velocity $\Omega$. We introduce a potential $\delta \Phi^{\rm S}$ that
corresponds to the absence of the body, and write the self-force 
difference as 
\begin{equation} 
\Delta \dot{\m} = -\m v^\alpha \nabla_\alpha 
\bigl( \delta \Phi - \delta \Phi^{\rm S} \bigr), 
\qquad 
\Delta \f_\alpha = \m \bigl( \delta_\alpha^{\ \beta} 
+ v_\alpha v^\beta \bigr) \nabla_\beta 
\bigl( \delta \Phi - \delta \Phi^{\rm S} \bigr).  
\label{sf_diff1}
\end{equation} 
Because $\delta \Phi$ and $\delta \Phi^{\rm S}$ are sourced by the
same particle on the same circular orbit, they are equally singular at
the position of the particle; their difference satisfies a homogeneous
wave equation and is smooth at $x=z(s)$. The actual self-force is then  
\begin{equation}  
\dot{\m} = \Delta \dot{\m} + \dot{\m}^{\rm S}, \qquad 
\f_\alpha = \Delta \f_\alpha + \f_\alpha^{\rm S}, 
\label{sf_full} 
\end{equation} 
with the second terms representing the self-force acting on the
particle in the absence of a fluid body. The equations governing 
$\delta\Phi^{\rm S}$ and the motion of the particle in this potential
are strictly identical to those governing the motion of a (constant)
scalar charge $q \equiv \m$ coupled to a scalar potential in flat
spacetime, and the resulting expression for the self-force is well
known (see, for example, Sec.~17.6 of
Ref.~\cite{poisson-pound-vega:11}). We have   
\begin{equation} 
\dot{\m}^{\rm S} = 0, \qquad 
\f_\alpha^{\rm S} = \frac{1}{3} \m^2    
\bigl( \eta_{\alpha\beta} + v_\alpha v_\beta \bigr) 
\frac{D^2 v^\beta}{ds^2}. 
\label{sf_S1} 
\end{equation} 
The only nonvanishing components of $\f_\alpha^{\rm S}$ are 
\begin{equation} 
\f_t^{\rm S} = -\Omega\, \f_\phi^{\rm S}, \qquad 
\f_\phi^{\rm S} = -\frac{1}{3} \m^2 \frac{M^{3/2}}{r_0^{5/2}} 
(1 + M/r_0) \qquad 
\label{sf_S2} 
\end{equation} 
for the circular orbit under consideration.  

The self-force difference can be evaluated straightforwardly by means 
of a (convergent) mode sum. We write 
\begin{equation} 
\delta \Phi^{\rm S} = \sum_{\ell m} P_{\ell m}^{\rm S}(r) 
Y_{\ell m}(\theta,\phi) e^{-im\Omega t} 
\end{equation} 
and let $\Delta P_{\ell m} := P_{\ell m} - P_{\ell m}^{\rm S}$. 
Inserting this and Eq.~(\ref{deltaPhi_out}) into
Eqs.~(\ref{sf_diff1}) and evaluating at $r=r_0$, $\theta=\pi/2$, and
$\phi = \Omega t$, we find after simple manipulations that   
\begin{subequations} 
\label{sf_diff2} 
\begin{align} 
\Delta \dot{\m} &= 0,  \\ 
\Delta \f_t &= -\Omega\, \Delta \f_\phi, \\ 
\Delta \f_r &= \m \sum_{\ell m} 
\frac{d}{dr} \Delta P_{\ell m}(r_0)\, Y_{\ell m}(\pi/2,0), \\ 
\Delta \f_\theta &= 0, \\ 
\Delta \f_\phi &= \m \sum_{\ell m} 
i m \Delta P_{\ell m}(r_0)\, Y_{\ell m}(\pi/2,0). 
\end{align} 
\end{subequations} 
Because $dP_{\ell m}/dr$ and $dP_{\ell m}^{\rm S}/dr$ are both
discontinuous at $r=r_0$, they must be consistently evaluated 
either at $r = r_0^+$ or at $r = r_0^-$. The choice is immaterial,
because the difference is continuous, but it is convenient to do the
evaluation at $r =r_0^+$. The mode sums can be simplified by
exploiting the reality conditions of Eq.~(\ref{P_reality}). By folding
the $m < 0$ part of the sum into the $m > 0$ part, we obtain 
\begin{subequations} 
\begin{align} 
\Delta \f_r &= \m \sum_\ell \biggl\{ 
\frac{d}{dr} \Delta P_{\ell 0}(r_0^+)\, Y_{\ell 0} (\pi/2,0)
+ 2 \sum_{m > 0} \Re  
\biggl[ \frac{d}{dr} \Delta P_{\ell m}(r_0) \biggr]  
Y_{\ell m} (\pi/2,0) \biggr\}, \\ 
\Delta \f_\phi &= -2\m \sum_\ell \sum_{m>0} 
m\, \Im \Bigl[ \Delta P_{\ell m}(r_0) \Bigr] Y_{\ell m}(\pi/2,0).
\end{align} 
\end{subequations} 

For the final expressions we import the results obtained in
Secs.~\ref{subsec:sol_general}, \ref{subsec:sol_m0}, and
\ref{subsec:sol_nostar} for the radial functions. After some simple
algebra we find that the self-force differences are given by     
\begin{equation} 
\Delta \dot{\m} = 0, \qquad
\Delta \f_t = -\Omega\, \Delta \f_\phi, \qquad
\Delta \f_r = \sum_{\ell=1}^\infty \bigl( \Delta \f_r \bigr)_\ell,
\qquad 
\Delta \f_\theta = 0, \qquad
\Delta \f_\phi = \sum_{\ell=1}^\infty \bigl( \Delta \f_\phi
\bigr)_\ell, 
\label{sf_diff3} 
\end{equation} 
with 
\begin{subequations} 
\label{sf_diff4} 
\begin{align} 
\bigl( \Delta \f_r \bigr)_\ell &= 
-\frac{\ell+1}{2\ell+1} \m G_{\ell 0} Y_{\ell 0} (\pi/2,0)
\frac{Z_\ell - 1}{r_0^2} 
\nonumber \\ & \quad \mbox{} 
+ 2 \sum_{m>0}^\ell \m G_{\ell m} Y_{\ell m} (\pi/2,0)\, \omega^2 
\Bigl\{ \alpha_\ell(\omega) j'_\ell(\omega r_0) 
- \bigl[ \beta_\ell(\omega) - j_\ell(\omega r_0) \bigr] 
y'_\ell(\omega r_0) \Bigr\}, \\ 
\bigl( \Delta \f_\phi \bigr)_\ell &= 
-2 \sum_{m>0}^\ell \m G_{\ell m} Y_{\ell m}(\pi/2,0)\, m\omega
\Bigl\{ \alpha_\ell(\omega) y_\ell(\omega r_0) 
+ \bigl[ \beta_\ell(\omega) - j_\ell(\omega r_0) \bigr] 
j_\ell(\omega r_0) \Bigr\}, 
\end{align} 
\end{subequations} 
where $\omega = m\Omega$. We observe that the sum over $\ell$ begins  
at $\ell = 1$; there is no contribution from $\ell = 0$ because as 
was first noticed in Sec.~\ref{subsec:sol_nostar}, $\Delta P_{\ell m}
= 0$ when $\ell = m = 0$. We recall that the sums over $m$ are
restricted to values such that $\ell + m$ is even; in this case we
find from Eqs.~(\ref{Glm}) and (\ref{Ylm}) that   
\begin{equation} 
\m G_{\ell m} Y_{\ell m}(\pi/2,0) = \frac{\m^2}{\gamma} 
(2\ell+1) \frac{(\ell-m-1)!!(\ell+m-1)!!}{(\ell-m)!! (\ell+m)!!}. 
\end{equation} 
The amplitudes $\alpha_\ell(\omega)$ and $\beta_\ell(\omega)$ must be
determined numerically by integrating the differential equations that
govern the perturbations of the fluid configuration; the shooting
method to achieve this was described in
Sec.~\ref{subsec:sol_general}. On the other hand, $Z_\ell-1$ was
obtained analytically in Sec.~\ref{subsec:sol_m0}, and its expression
is displayed in Eq.~(\ref{Z_expression}).   

The complete self-force is constructed by combining
Eqs.~(\ref{sf_full}), (\ref{sf_S2}), (\ref{sf_diff3}),
(\ref{sf_diff4}) and performing the mode sums. An outcome of this 
computation is that $\dot{\m} = 0$ --- the particle's mass is actually
constant when the particle moves on a circular orbit. Our results for
the self-force were already displayed in Fig.~\ref{fig:fig2}, and we
provide additional details in a number of additional figures. 
In all of them, the compactness of the fluid body is fixed to
$\chi = 0.3$. In Fig.~\ref{fig:fig8} we examine the contributions
to $\Delta \f_r$ and $\Delta \f_\phi$ coming from the lowest multipole
orders $\ell$ in the mode sum, for selected values of $r_0$. We see that
when $r_0$ is small, the convergence of the mode sum for $\Delta \f_r$ 
is slowed down by the resonant features, which can produce abnormally 
large contributions for selected values of $\ell$. We see also that the
convergence of the mode sum for $\Delta \f_\phi$ is considerably faster, 
and much less affected by the resonant features. In Fig.~\ref{fig:fig9},
we display $\f_r$ and $\f_\phi$ in small intervals of $r_0$, and
associate each resonant feature with a specific quasinormal mode. 
We see that the broad resonance observed in Sec.~\ref{sec:intro} 
is associated with the $\ell=2$ $f$-mode of the fluid-gravity system,
and that the resonance is triggered when $m=2$, so that 
$2\Omega = \Re(\omega)$. Finally, note that only resonances with 
low-$\ell$ quasinormal modes have been clearly identified 
in the figures. In fact, as $\ell$ increases, the modes become 
longer lived and the resonances get accordingly narrower. 
To produce Figs.~\ref{fig:fig2} and \ref{fig:fig9} we sampled more 
densely the regions where resonances with the computed low-multipole
quasinormal modes were expected to arise, and many of the features
displayed in these figures could only be identified through this more 
refined search.

\subsection{Orbital evolution} 

The self-force computed in the preceding section drives an evolution
of the circular orbit: the orbital radius $r_0$ will slowly decrease
because of the dissipative action of the force, and the orbital
frequency $\Omega$ will also change because of both dissipative and 
conservative effects. In this subsection we derive and integrate the
equations that govern this orbital evolution, exploiting a two-time
expansion initially formulated by Pound \cite{pound:15}.  

We write the orbital equation as 
\begin{equation} 
\frac{D v^\alpha}{ds} = \bigl( \eta^{\alpha\beta} + v^\alpha v^\beta
\bigr) \nabla_\beta \Phi + F^\alpha,
\end{equation} 
in which $v^\alpha = dz^\alpha/ds$ is the four-velocity, $s$ is proper
time, $\Phi = M/r$ is the background gravitational potential, and
$F^\alpha := \m^{-1} \f^\alpha$ is the self-force per unit mass. For
our purposes it is convenient to adopt the coordinate time $t$ as
orbital parameter (instead of proper time $s$), and in this notation
the orbital equation becomes 
\begin{equation} 
0 = Z^\alpha := \ddot{z}^\alpha 
+ \Gamma^\alpha_{\beta\gamma} \dot{z}^\beta \dot{z}^\gamma 
+ \frac{\dot{\gamma}}{\gamma} \dot{z}^\alpha 
- \frac{1}{\gamma^2} \Bigl[ \bigl( g^{\alpha\beta} 
+ \gamma^2 \dot{z}^\alpha \dot{z}^\beta \bigr) \nabla_\beta \Phi 
+ F^\alpha \Bigr], 
\end{equation} 
where an overdot now indicates differentiation with respect to $t$,
and $\gamma := dt/ds$ is given by  
\begin{equation} 
\frac{1}{\gamma^2} = -\eta_{\alpha\beta} \dot{z}^\alpha \dot{z}^\beta. 
\end{equation} 

We assume that the particle follows a circular orbit with
slowly-changing orbital parameters $r_0$ and $\Omega$. To express this
mathematically we introduce a slow-time variable 
$\bar{t} := \epsilon t$, where $\epsilon \ll 1$, and write 
\begin{equation} 
r_0 = r_0(\bar{t}), \qquad 
\Omega = \Omega(\bar{t}). 
\end{equation} 
The orbital phase $\phi$ is then obtained by integrating the orbital
frequency with respect to (fast) time: 
\begin{equation} 
\phi = \int \Omega(\bar{t})\, dt 
= \frac{1}{\epsilon} \int \Omega(\bar{t})\, d\bar{t}. 
\end{equation} 
With this we find that the vector tangent to the orbit can be
expressed as 
\begin{equation} 
\dot{z}^\alpha = \bigl( 1, \epsilon r_0'(\bar{t}), 0, \Omega(\bar{t})
\bigr), 
\end{equation} 
in which a prime indicates differentiation with respect to
$\bar{t}$. We also have 
\begin{equation} 
\ddot{z}^\alpha = \bigl( 0,  O(\epsilon^2), 0, \epsilon
\Omega'(\bar{t}) \bigr). 
\end{equation} 
With these ingredients we have that $\gamma^{-2} = 1 - r_0^2 \Omega^2
+ O(\epsilon^2)$, $\dot{\gamma}/\gamma = \epsilon \gamma^2( r_0 \Omega^2
r_0' + r_0^2 \Omega \Omega')$, and we further expand $\Omega$ and
$F^\alpha$ in powers of $\epsilon$: 
\begin{equation} 
\Omega(\bar{t}) = \Omega_0(\bar{t}) + \epsilon \Omega_1(\bar{t}) 
+ O(\epsilon^2), \qquad 
F^\alpha = \epsilon F_1^\alpha\bigl( r_0(\bar{t}) \bigr) 
+ O(\epsilon^2).  
\end{equation} 
The second equation incorporates an assumption that to leading order
in $\epsilon$, the self-force depends only on the orbital radius
$r_0$. 

The foregoing relations can all be inserted into the orbital
equations $Z^\alpha = 0$, which are then expanded in powers of
$\epsilon$. At order $\epsilon^0$, $Z^r = 0$ produces 
\begin{equation} 
\Omega_0^2 =\frac{M/r_0^3}{1+M/r_0}, 
\end{equation} 
the same expression as in Eq.~(\ref{v_def}). At order $\epsilon$,
$Z^r=0$ gives instead 
\begin{equation} 
\Omega_1 = -\frac{1}{2} (r_0/M)^{1/2}(1+M/r_0)^{-3/2} F_{1\, r}, 
\end{equation} 
while $Z^\phi = 0$ yields 
\begin{equation} 
r_0' = 2 (r_0/M)^{1/2}(1+M/r_0)^{-1/2}(1+2M/r_0)^{-1} F_{1\, \phi}
+ O(\epsilon). 
\end{equation} 
These equations provide a complete account of the orbital
evolution. Reverting to the original notation, we have that 
\begin{subequations} 
\begin{align} 
\dot{r}_0 &= 2 (r_0/M)^{1/2}(1+M/r_0)^{-1/2}(1+2M/r_0)^{-1} 
F_{\phi}(r_0) + O(\epsilon^2), \\ 
\Omega &= \sqrt{ \frac{M/r_0^3}{1+M/r_0} } 
-\frac{1}{2} (r_0/M)^{1/2}(1+M/r_0)^{-3/2} F_{r}(r_0)
+ O(\epsilon^2), \\ 
\phi &= \int \frac{\Omega}{\dot{r}_0}\, dr_0. 
\end{align} 
\end{subequations}   
The equations reveal that $F_\phi := \m^{-1} \f_\phi$, the dissipative  
component of the self-force, determines the evolution of the orbital
radius $r_0$, and therefore the first term in the expression for the
angular velocity $\Omega$. They show also that $F_r := \m^{-1} \f_r$,
the conservative component of the self-force, produces an additional
shift in the angular velocity.  

\begin{figure}
\includegraphics[width=0.9\linewidth]{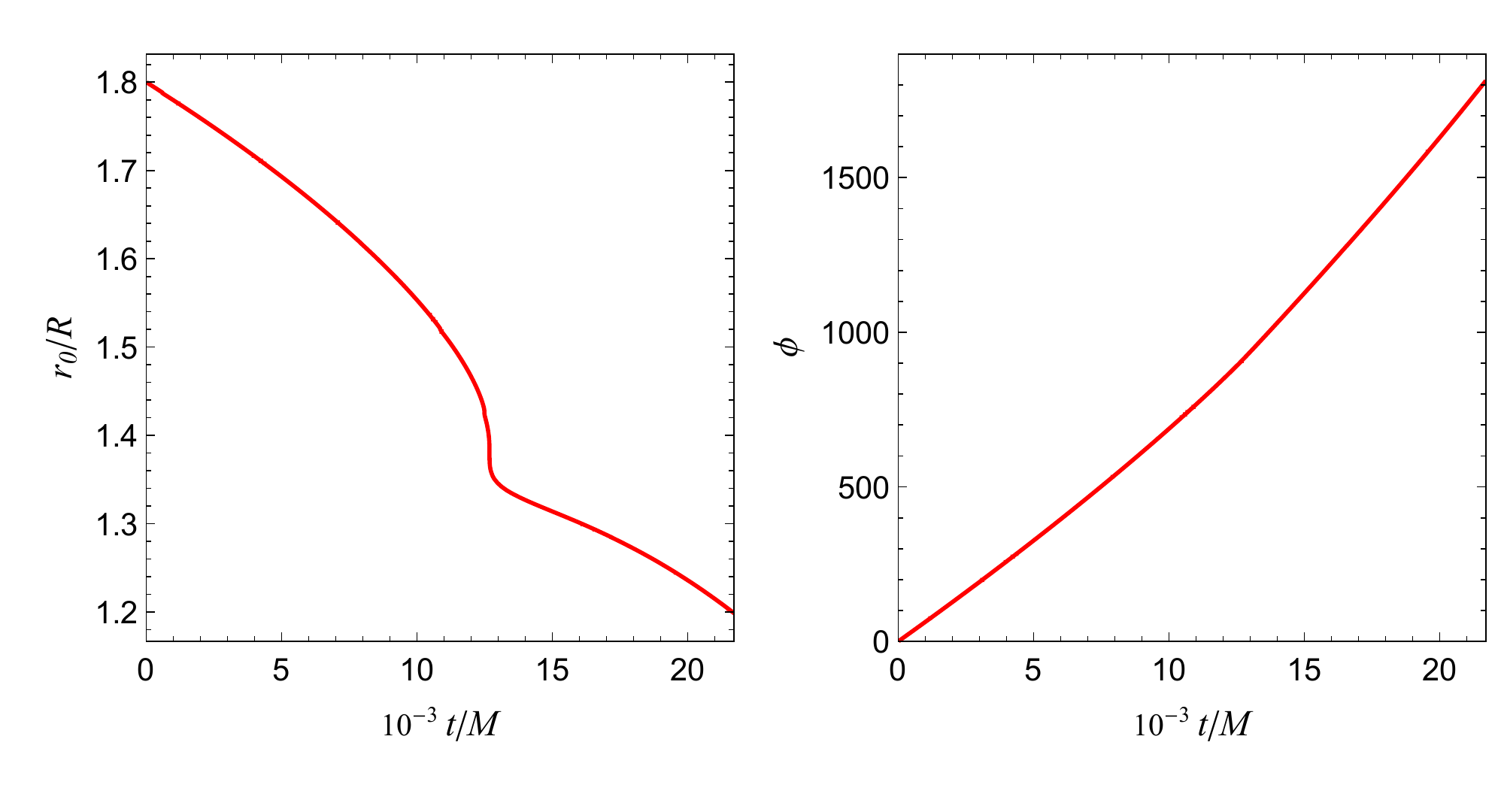}
\caption{Plot of the orbital radius $r_0/R$ and the orbital phase
  $\phi$ as functions of $t/M$. The calculations were performed with
  $\m/M = 0.01$.} 
\label{fig:fig12} 
\end{figure} 

We next insert the self-force data obtained previously into these
equations, and integrate them numerically to obtain $r_0(t)$,
$\Omega(t)$, and $\phi(t)$. For concreteness, we set the mass 
of the particle to be $\m= 0.01 M$, and compute its evolution from an 
initial orbit with $r_0 = 1.80$ to a final orbit with $r_0=1.20$.
Note that a rescaling of the mass would result in a corresponding
rescaling of the inspiral time scale, and would also affect the shift 
$\Delta \Omega := \Omega - \Omega_0$ of the angular frequency relative
to the purely dissipative value. The most revealing information 
regarding both dissipative and conservative aspects of the orbital
evolution is provided by $\Omega(t)$, which was already displayed in
Fig.~\ref{fig:fig3} and discussed in Sec.~\ref{sec:intro}. In
Fig.~\ref{fig:fig12} we also provide plots of $r_0(t)$ and
$\phi(t)$. Because $r_0(t)$ is calculated entirely from the
dissipative component of the self-force, it is relatively featureless
compared with $\Omega(t)$; an exception is the rapid decrease when
$t/M \simeq 12,680$, which is caused by the broad resonance near
$r_0/R \simeq 1.38$. The orbital phase is also relatively featureless,
in spite of the large excursions of the angular velocity produced by
the conservative component to the self-force.
 
\begin{acknowledgments} 
One of us (EP) is beholden to the Canadian Institute of Theoretical 
Astrophysics for its warm hospitality during a research leave from the 
University of Guelph. This work was supported by the Natural Sciences
and Engineering Research Council of Canada, the Conselho Nacional de
Desenvolvimento Cient\'{\i}fico e Tecnol\'ogico (CNPq), and the JSPS Postdoctoral Fellowships for Research Abroad.       
\end{acknowledgments}    

\bibliography{../bib/master} 

\end{document}